\documentclass[11pt,a4paper]{article} 
\pdfoutput=1
\usepackage{jheppub}                  

\usepackage{graphicx}

\usepackage[normalem]{ulem}	

\newcommand{\nova}{NO$\nu$A~}
\newcommand{\dcp}{\delta_{\rm CP}}

\def\lsim{\raise0.3ex\hbox{$\;<$\kern-0.75em\raise-1.1ex
\hbox{$\sim\;$}}}
\def\gsim{\raise0.3ex\hbox{$\;>$\kern-0.75em\raise-1.1ex
\hbox{$\sim\;$}}}

\title{
What can we learn about the lepton CP phase in the next 10 years? 
}
\author{P.~A.~N.~Machado$^{1}$}
\author{H.~Minakata$^{2}$}
\author{H.~Nunokawa$^{2}$}
\author{R.~Zukanovich Funchal$^{1}$}
\affiliation{
$^1$Instituto de F\'{\i}sica, Universidade de S\~ao Paulo, C.\ P.\
66.318, 05315-970 S\~ao Paulo, Brazil \\
$^2$Departamento de F\'{\i}sica, Pontif{\'\i}cia Universidade Cat{\'o}lica 
do Rio de Janeiro, C. P. 38071, 22452-970, Rio de Janeiro, Brazil  \\ 
}

\date{\today}

\abstract{ We discuss how the lepton CP phase can be constrained by
  accelerator and reactor measurements in an era without dedicated
  experiments for CP violation search.  To characterize globally the
  sensitivity to the CP phase $\delta_{\rm CP}$, we use the CP
  exclusion fraction, which quantifies what fraction of the
  $\delta_{\rm CP}$ space can be excluded at given input
  values of $\theta_{23}$ and $\delta_{\rm CP}$. Using the measure we
  study the CP sensitivity which may be possessed by the accelerator
  experiments T2K and NO$\nu$A. We show that, if the mass hierarchy is
  known, T2K and NO$\nu$A alone may exclude, respectively, about
  $50\%-60\%$ and $40\%-50\%$ of the $\delta_{\rm CP}$ space at 90\%
  CL by 10 years running, provided that a considerable fraction of
  beam time is devoted to the antineutrino run.  The synergy between
  T2K and NO$\nu$A is remarkable, leading to the determination of the
  mass hierarchy through CP sensitivity at the same CL.  }


\emailAdd{accioly@if.usp.br}
\emailAdd{hisakazu.minakata@gmail.com}
\emailAdd{nunokawa@puc-rio.br}
\emailAdd{zukanov@if.usp.br}

\begin{document} 

\maketitle

\section{Introduction}
\label{introduction}

After accumulating hints and indications, the elusive lepton mixing
angle $\theta_{13}$ was finally discovered to be non-zero and measured
with high
precision~\cite{Abe:2011sj,Abe:2013xua,Adamson:2011qu,Adamson:2013ue,Abe:2011fz,Abe:2012tg,An:2012eh,An:2012bu,Ahn:2012nd}.
Thus, we are left with the CP violating phase $\delta_{\rm CP}$, the
unique unknown parameter in the lepton flavor mixing matrix
\cite{Maki:1962mu}, which could remain  a mystery for sometime
together with the problem of determining the neutrino mass
hierarchy. Lepton CP violation due to $\delta_{\rm CP}$, in
association with the one by the possible Majorana phases, may hide the
secret behind the baryon number asymmetry in our universe
\cite{leptogenesis}. However, because of the smallness of the effects
of $\delta_{\rm CP}$, being suppressed by the small ratio of two
$\Delta m^2$ and products of mixing angles, its measurement will
require dedicated facilities such as Hyper-Kamiokande~\cite{Hyper-K}
and LBNE~\cite{LBNE} as well as intense neutrino beams.

Here, the potential problem is that it will take a long time, $\sim
10$ years, to construct and operate such facilities. Therefore, it may
be worthwhile to ask the question, ``What can be done in the next 10
years toward the observation of lepton CP violation?''.  To sharpen up
our concern we may ask a more scrutinizing question: ``How can an
experiment that is not actually capable of observing CP violation due
to $\delta_{\rm CP}$ help us to pave the way to the final
discovery?''. It is the purpose of this paper to give a partial answer
to these questions.

To reveal the sensitivity to CP violation at a particular time, one
can think of two different approaches: bring all available data
together to collect every tiny piece of information on $\delta_{\rm
  CP}$ in them to enhance CP sensitivity, the spirit of the so called
global fits \cite{Tortola:2012te,Fogli:2012ua,GonzalezGarcia:2012sz},
or focus on a few measurements which have relatively higher
sensitivities to CP. In this paper, following our previous analysis
\cite{Machado:2011ar}, we take the latter strategy with implementing
the precision reactor measurement of $\theta_{13}$
\cite{Minakata:2003wq}. There are pros and cons in each approach. In a
global fit the sensitivity is higher, but it is achieved at the price
of combining many experiments with different systematic errors. In our
approach that drawback is somewhat cured though it may not reveal the
best possible sensitivity to CP violation. We believe that it is
important to proceed in both ways as they are complementary to each
other.

In addressing CP violation, one of the relevant issues is how the CP
sensitivity achievable by a particular experimental setting can be
quantified and displayed. Though it might sound a bit technical, this
is an important point because getting a robust hint, although it could
still be only a slight indication, is an important step for the
successful completion of the long-term race toward detecting and
measuring the lepton CP violating phase $\delta_{\rm CP}$, {\em the
  marathon in neutrino physics}. To quantify the maximum sensitivity
possessed by a particular experiment, or by a set of experiments to
measure $\delta_{\rm CP}$, we analyze the fraction of values of
$\delta_{\rm CP}$ that can be excluded for a given set of input
parameters, which we call the ``CP exclusion fraction''.
Notice that it is essentially the same measure as the ``CP coverage"
which was introduced in \cite{Winter:2003ye} and extensively used in
the analyses in \cite{Huber:2004gg}. 
See Sec.~\ref{CP-ex-fraction} and Appendix \ref{fCPX} 
for more about the relationship between the measures
for CP sensitivity.

We argue that one of the most important goals related to lepton CP
violation that may be reached by the ongoing and the upcoming
experiments is to exclude a significant fraction of the $\delta_{\rm
  CP}$ space. The CP exclusion fraction will serve for the discussion
of this point. We will use the global measure to investigate the CP
exclusion potential of T2K and NO$\nu$A within a 10 years
perspective. It is interesting and timely to discuss the following
questions: What is the impact of running T2K also in the antineutrino
mode on the determination of $\delta_{\rm CP}$? What would be the
optimal time sharing between neutrino and antineutrino beams in order
that T2K can say something meaningful on $\delta_{\rm CP}$? How T2K
and NO$\nu$A compare with each other in $\delta_{\rm CP}$ sensitivity?
Can the combination of equal-time running of T2K and NO$\nu$A say more
on $\delta_{\rm CP}$ than each one of these experiments with doubled
running time? Or, rephrasing, is there a synergy between them?

Our results demonstrates that CP sensitivities which may be achievable
by 10 years running of T2K and NO$\nu$A are not so low, even after
admitting the fact that these experiments are not originally designed
to discover CP violation.  We have found that running T2K in the
antineutrino mode makes the experiment, in general, much more powerful
in excluding regions of $\delta_{\rm CP}$ in a way independent of the
neutrino mass hierarchy, the $\theta_{23}$ octant, and of the sign of
$\sin\delta_{\rm CP}$. Our study shows that the optimal setting would
be to run about half the time in neutrino and the other half in
antineutrino mode. If we compare these two experiments, it appears
that T2K has better sensitivity for CP, but NO$\nu$A can make an
unique contribution by its higher sensitivity to the matter
effects. As a consequence the synergy between these two experiments is
quite visible. See, for example, \cite{Huber:2009xx,Ghosh:2013yon} for
related works.

\section{CP exclusion fraction; A measure of CP sensitivity for non-conclusive experiments}
\label{CP-ex-fraction}

In this paper, we investigate the experimental sensitivity to
$\delta_{\rm CP}$ of T2K and NO$\nu$A and quantify it by using the
fraction of $\delta_{\rm CP}$ values which can be disfavored by these
experiments for a given set of input parameters.  We call this
fraction the ``CP exclusion fraction'' $\equiv f_{\rm CPX}$. As
explained in more detail in Appendix~\ref{fCPX-def}, $f_{\rm CPX}$ is
calculated as the fraction of $\delta_{\rm CP}$ $\in [-\pi,\pi]$
values which can be excluded by the experiment at a given confidence
level for each input point of the parameter space $(\sin^2
\theta_{23}^{\rm in}, \delta_{\rm CP}^{\rm in})$. It is thus a global
measure which covers the entire input parameter space.
It should be mentioned that the CP exclusion fraction is related to
the ``CP coverage'' defined and extensively used in
\cite{Winter:2003ye,Huber:2004gg} as $f_{\rm CPX}=1- {\rm
  CP~coverage}/360^{\circ}$. Instead of using the CP coverage we choose to
work with CP exclusion fraction for an appeal to intuition to
facilitate understanding the plots.
In this work, we use the standard parameterization for the neutrino
mixing angles as well as the CP phase $\delta_{\text{CP}}$ found in
Ref.~\cite{Beringer:1900zz}.

While expert readers can go directly to Sec.~\ref{CP-T2K} for the
analysis results, a comparative discussion of CP exclusion fraction
$f_{\rm CPX}$ with CP violation (CPV) fraction may be illuminating for
a wide range of non-expert readers to reveal the nature of the two
measures for CP sensitivity and their difference. The latter gives us
the fraction of $\delta_{\rm CP}$ values for which CPV can be
established as a function of the input parameter values, usually as a
function of $\sin^ 2\theta_{13}$.\footnote{
The CPV fraction is used in many papers including, for example,
Refs.~\cite{Hyper-K,Huber:2009xx}.  }
We first note that in the general context of revealing CP sensitivity,
they are complementary to each other. Then, what are the differences?

The CP exclusion fraction plot is a particularly useful tool to reveal
the potential for exploring the CP phase effects by a ``non-conclusive
experiment'' which is not designed as a dedicated CP violation
discoverer.  Suppose that there are two experiments each of which
alone can not discover (establish) CPV at a given CL. In this case the
CPV fraction vanishes for both experiments, and it does not provide us
with any useful informations. But, with the use of $f_{\rm CPX}$ we are
able to reveal the CP sensitivity of each experiment and can tell which
one has higher capability of restricting the allowed range of
$\delta_{\text{CP}}$. In this way, the CP exclusion fraction serves as
a viable way of quantifying the experimental CP sensitivity for
non-conclusive experiments, and provides a better chance for a fruitful
discussion of synergy.

The merit of using the CPV fraction is that it conveys a clear cut
message by focusing on ``yes or no'' to CP violation. Because of the
definition, however, it suffers from the ``bias'' of choosing
$\delta_{\rm CP}$ equal 0 or $\pi$ as a reference point to measure the
capability of detecting CP violation. That is, the CPV fraction plot
neither tells us whether the experiment is able to exclude for
example, $\delta_{\rm CP} = {\pi}/{2}$ or $-{\pi}/{2}$, nor allows us
to extract the precision on $\delta_{\rm CP}$ determination, {\it
  e.g.}, at around these points. We emphasize that the exclusion of
the region around $\delta_{\rm CP} = \pm {\pi}/{2}$, depending upon
the mass hierarchy, is likely to be the initial footprint of the near
future experiments which first step into probing the CP phase.

\section{Sensitivity to CP phase expected by T2K}
\label{CP-T2K}

In this and the following sections we discuss the results of our
analyses, the sensitivities to CP phase determination or exclusion to
be expected by the T2K and NO$\nu$A experiments, respectively,
assuming accurate measurement of $\theta_{13}$ by the reactor
experiments. Details of our analysis method are described in
Appendix~\ref{method}. An intuitive understanding of some of the salient
features of the analysis results will be offered in Appendix~\ref{sec:bi-p}.

Considering the nature of the experiments as the initial phase of CP
measurement we will use, throughout this section, the CP exclusion
fraction in $\delta_{\text{CP}} - \sin^2 \theta_{23}$ space defined at
90\% CL to display the sensitivity to CP phase
$\delta_{\text{CP}}$.\footnote{
Of course, since $\theta_{13}$ has been measured rather accurately it
is now more appropriate to discuss the sensitivity to $\delta_{\rm
  CP}$ in the $\delta_{\rm CP} ~{\rm vs}~\sin^2 \theta_{23}$ space as
$\theta_{23}$ is now the least known angle.  }
%
We note that while 90\% CL may not guarantee high enough confidence
for exclusion, the criterion is often used to place useful constraints
on physics parameters in the literatures, for example, in the reports
from Bugey \cite{Declais:1994su}, Chooz~\cite{Apollonio:1999ae}, and
T2K~\cite{Abe:2011sj} experiments.
While we show only the results corresponding to 90\% CL in this paper,
we have also performed the computations to obtain the contours at 95\%
CL ($\simeq 2 \sigma$ CL).  Very roughly speaking, the change of CP
exclusion fraction when we use 95\% CL is that the contours of equal
$f_{\text{CPX}}$ at 90\% CL are to be interpreted as $f_{\text{CPX}} -
(0.1 - 0.15)$ at 95\% CL, the precise values of $f_{\text{CPX}}$
reduction depend on $\delta_{\text{CP}}$ and $\sin^2 \theta_{23}$.

We focus our discussion primarily on the possibility of a total of 10
years of data taking. The reason being, as we will see shortly, that
after a total of 5 running years T2K will only be able to exclude 50\%
of $\delta_{\text{CP}}$ values in a very limited parameter space in
the $\delta_{\text{CP}} - \sin^2 \theta_{23}$ plane, even if we assume
that the mass hierarchy is known. We would like to explore the
possibility of increasing the CP sensitivity of the experiment in a
longer time span. As we mentioned in Sec.~\ref{introduction}, most
probably, the construction of a dedicated CP explorer needs longer
than 10 years from now, so that it is not an unrealistic scenario to
examine.

The inverted mass hierarchy has been favored by some experimental
analyses~\cite{Itow:2013zza,MINOS-full}, however feebly. Hence, the
choice of the hierarchy to be displayed in our figures is basically
arbitrary, and we opt for the inverted one.
Our treatment will not be completely equal for T2K and NO$\nu$A,
because our analysis of NO$\nu$A can not be as mature as that of T2K
for which we can profit from the informations of the experiment in
operation.

\subsection{Total of 5 running years ($5\times 10^{21}~{\rm POT}$)}
\label{5years}

\begin{figure}[htbp]
\begin{center}
\vspace{2mm}
\includegraphics[width=0.31\textwidth]{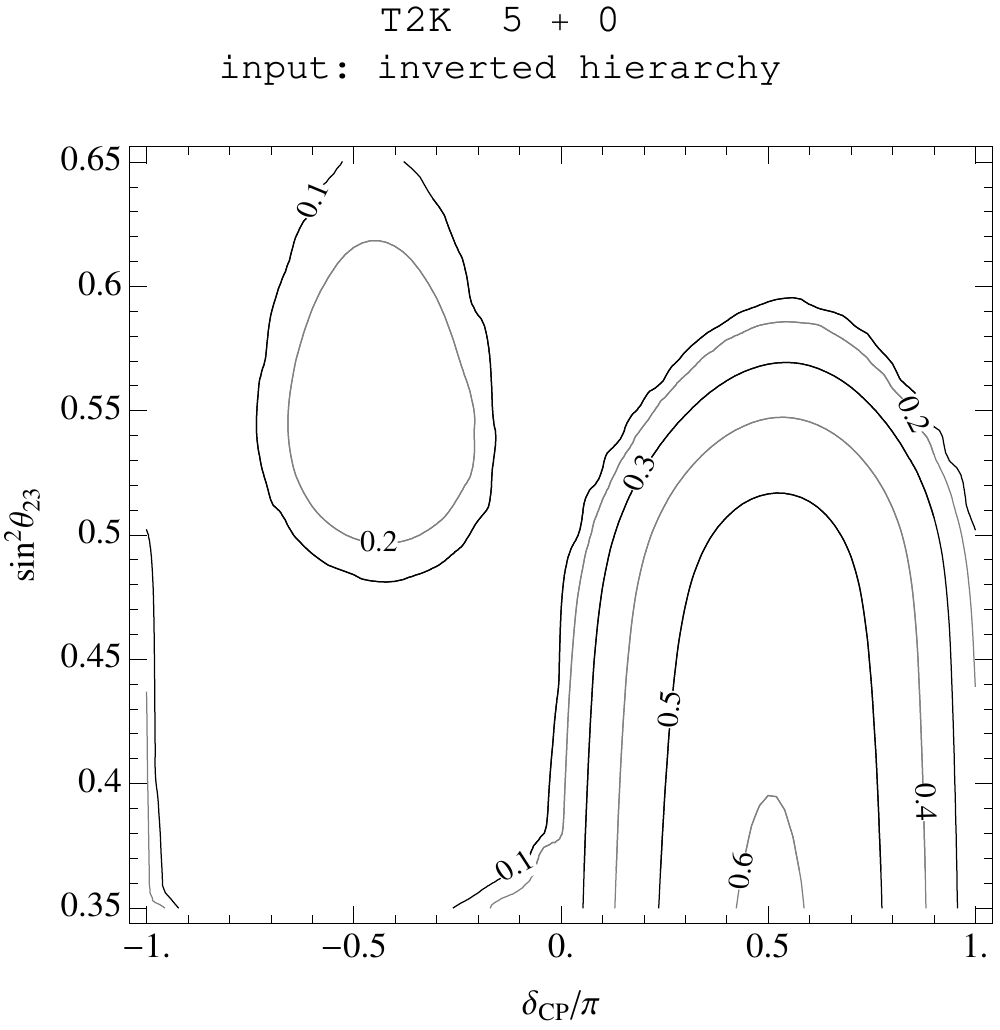}
\includegraphics[width=0.31\textwidth]{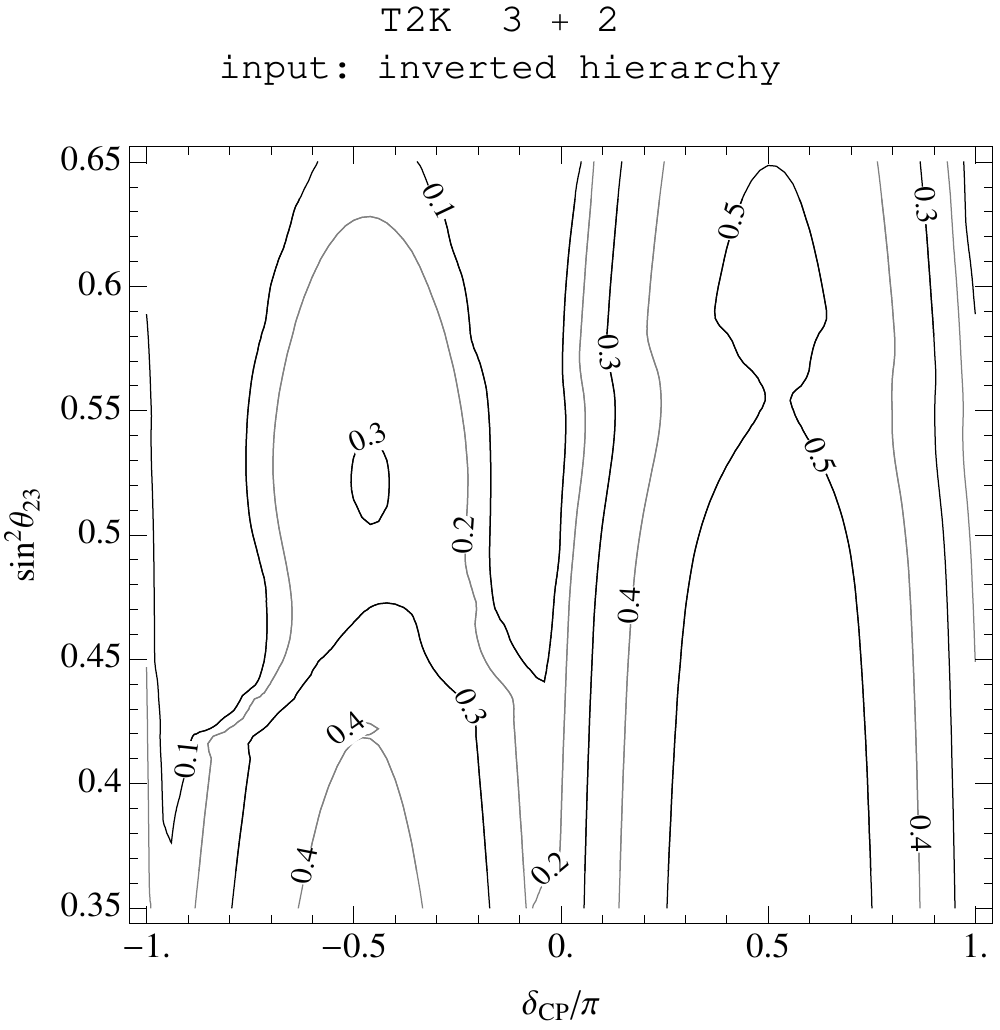}
\includegraphics[width=0.31\textwidth]{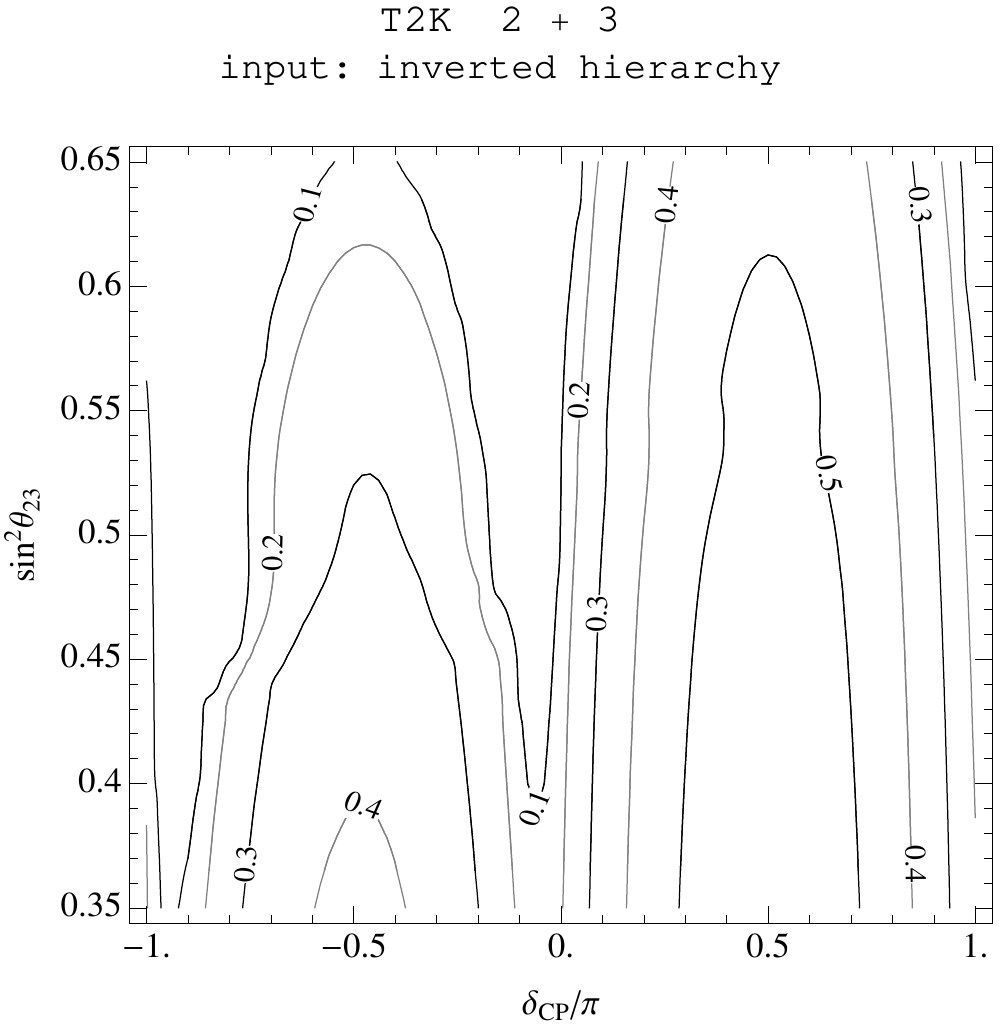}
\includegraphics[width=0.31\textwidth]{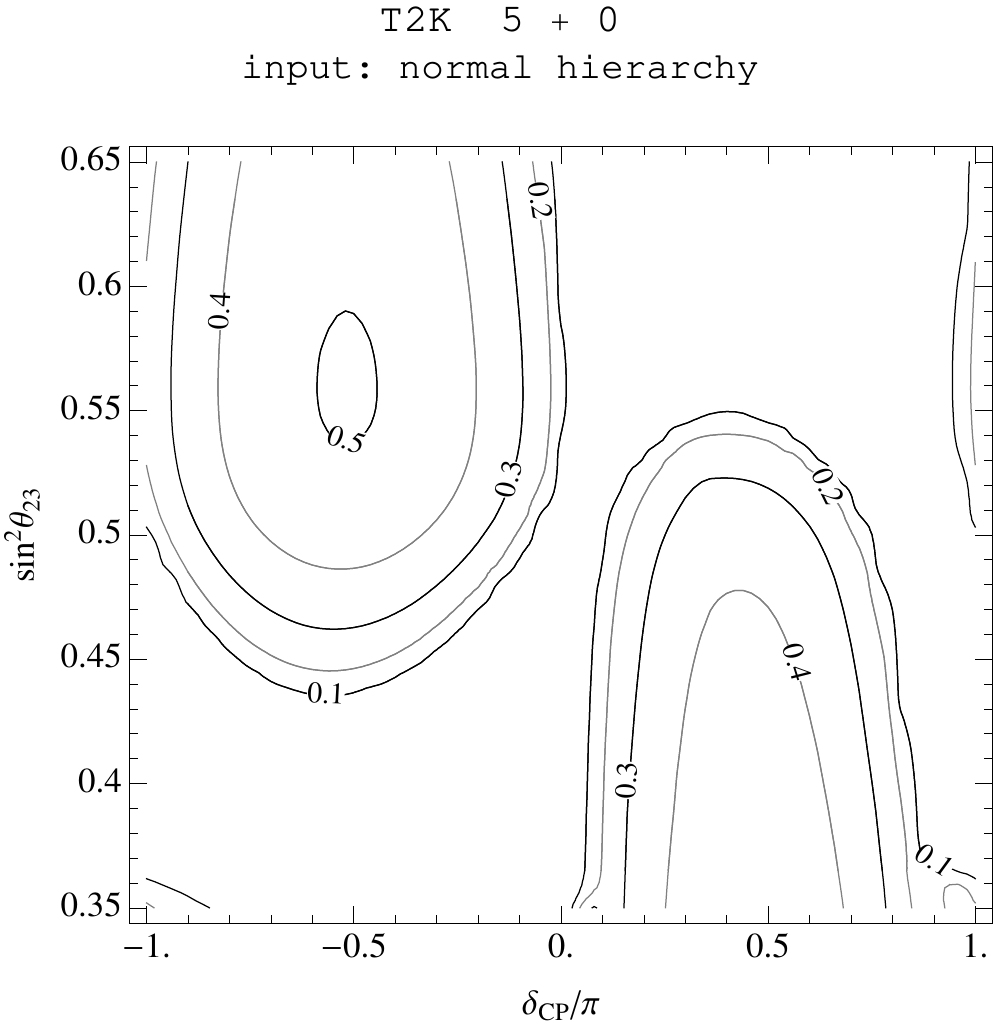}
\includegraphics[width=0.31\textwidth]{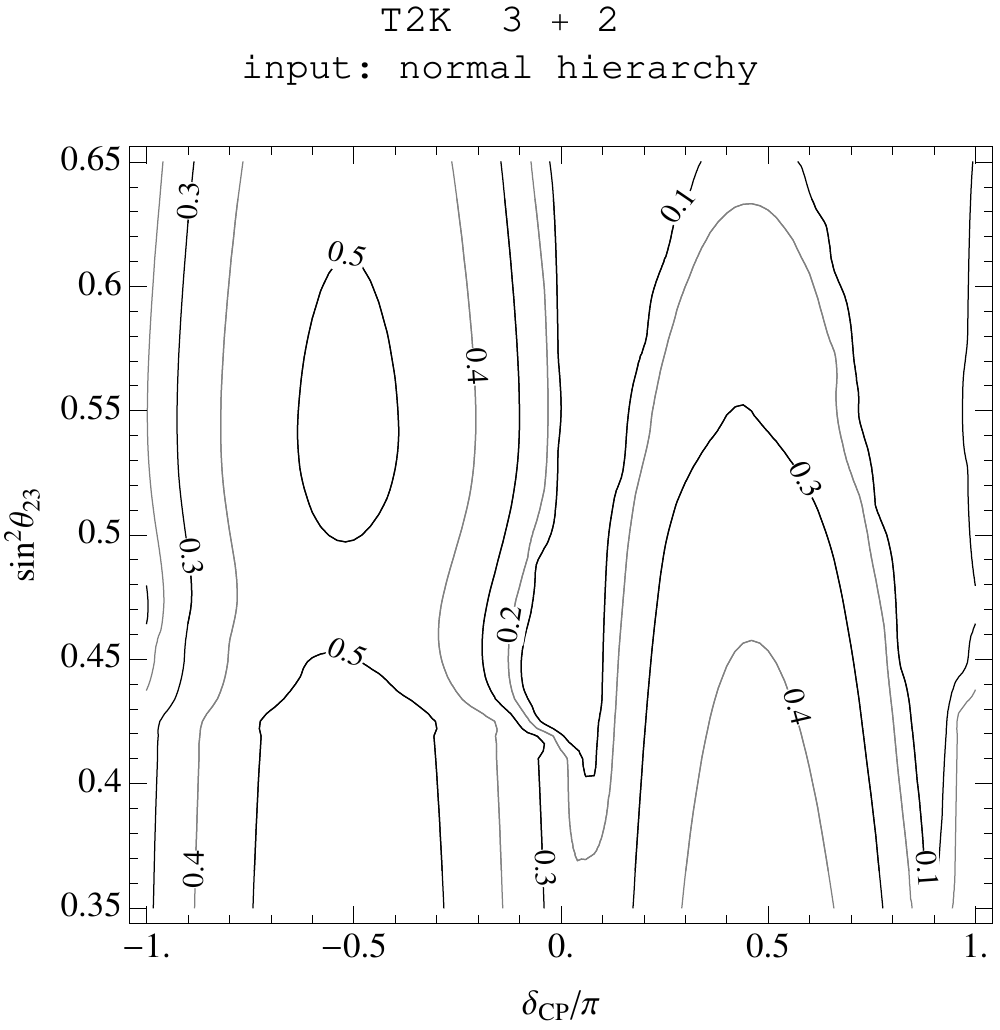}
\includegraphics[width=0.31\textwidth]{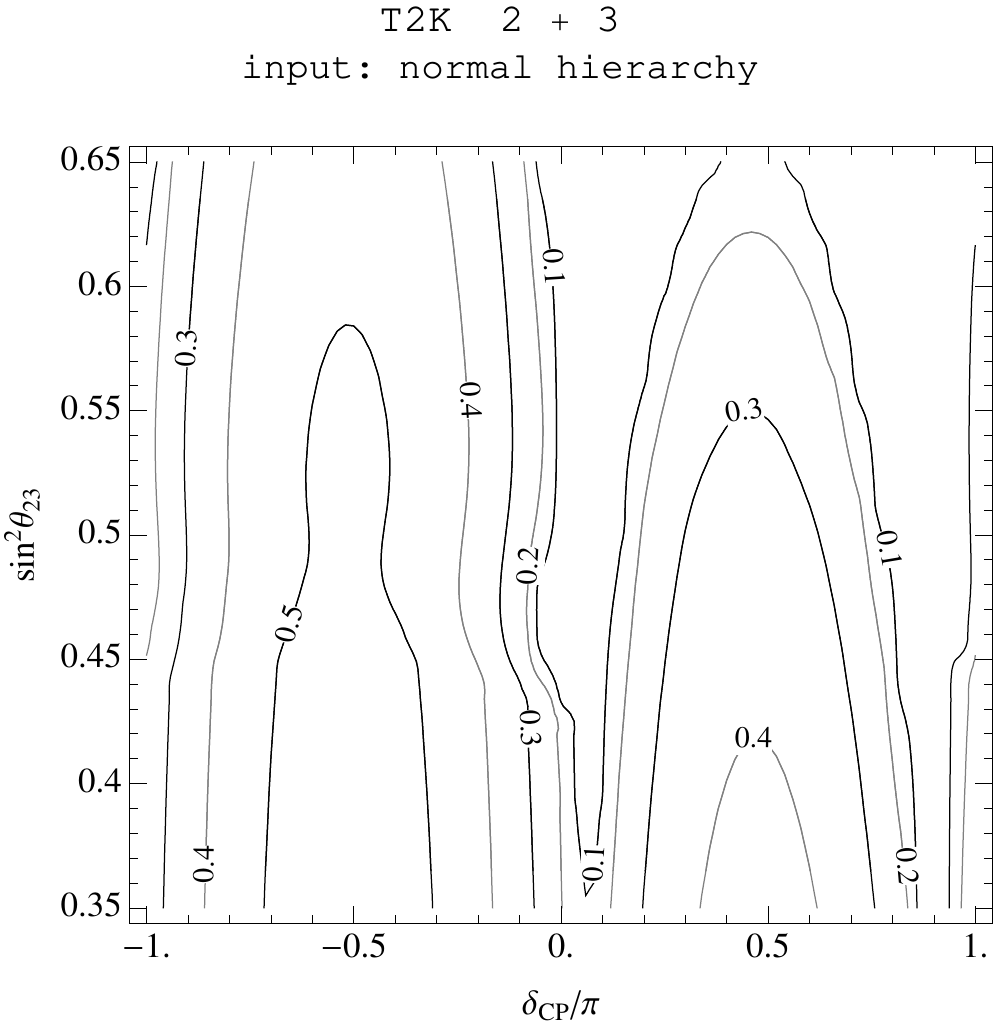}
\end{center}
\vspace{-6mm}
\caption{CP exclusion fraction isolines plotted on the
  $\delta_{\text{CP}} - \sin^2 \theta_{23}$ plane at 90 \% CL, for T2K
  running in $\nu$+$\bar \nu$ mode for $5+0$ (left), $3+2$ (center)
  and $2+3$ (right) years. The top (bottom) panels are for the case of
  inverted (normal) input mass hierarchy. The fit marginalizes over
  both hierarchies.}
\label{T2K-5}
\end{figure}

In Fig.~\ref{T2K-5}, the contours of equal CP exclusion fraction are plotted in
the space spanned by the true values of $\delta_{\text{CP}}$ and
$\sin^2\theta_{23}$. A total running time of 5 years is assumed with the nominal
design luminosity, and the results for the $\nu + \bar{\nu}$ beam time sharing
of $5+0$, $3+2$, and $2+3$ years are shown (panels from left to
right). Intermediate runnings, like $4+1$ years, lie between the results
shown. In the upper panels (lower panels) of Fig.~\ref{T2K-5} the inverted
(normal) hierarchy is assumed as the input true mass hierarchy. It is quite
likely that the mass hierarchy will not be determined with high confidence level
when T2K completes its running period of 5 years. Therefore, we present here
only the case where we fit for an unknown mass hierarchy, obtained by
marginalizing over both cases.

The numbers on the isolines correspond to the CP exclusion
  fraction that can be achieved at 90\% CL. By comparing the CP
exclusion fractions of the three cases of $\nu+\bar{\nu}$ running
periods of $5+0$, $3+2$, and $2+3$ years in Fig.~\ref{T2K-5}, it is
evident that running in antineutrino mode helps to improve the CP
sensitivity. 
It is notable that the performance of $3+2$
and $2+3$ years of runnings are roughly comparable to each other.

We note some characteristic features of the exclusion
fraction iso-contour lines we can see in Fig.~\ref{T2K-5}:

\begin{itemize}

\item Overall, the regions of relatively high sensitivity to CP are
  centered around $\delta_{\text{CP}} \simeq \pm {\pi}/{2}$.

\item 
In the $5+0$ years running option the CP sensitive region is
restricted mostly to two regions centered at ($\delta_{\text{CP}}
\simeq {\pi}/{2}$, low $\sin^2\theta_{23}$) and ($\delta_{\text{CP}}
\simeq - {\pi}/{2}$, high $\sin^2\theta_{23}$), whereas in $3+2$ and
$2+3$ years running options (center and right panels) the dependence
on $\sin^2\theta_{23}$ is weakened, particularly at around
$\delta_{\text{CP}} \simeq {\pi}/{2}$ and $\delta_{\text{CP}} \simeq -
{\pi}/{2}$ for the inverted and the normal hierarchies, respectively.

\end{itemize}
\noindent

From the probability point of view, one naively expects that the
highest sensitivity to CP would be at around $\delta_{\text{CP}}
\simeq \pm {\pi}/{2}$, in agreement with the first feature mentioned
above. However, as statistics increases these most favorable values
become less favorable than $\delta_{\text{CP}}=0$, depending on the
$\theta_{23}$ value and our knowledge on the mass hierarchy,
as will be shown in
Figs.~\ref{T2K-10-IH}-\ref{T2K-NOVA}.
An attempt to explain such a change in behavior in terms of the
bi-probability plot can be found in Appendix~\ref{sec:bi-p} (see
Fig.~\ref{Bi-P-plots-2} and its description).  The second feature
explained above regarding the dependence on $\theta_{23}$ can also be
understood qualitatively in terms of the bi-probability plot, see
Fig.~\ref{Bi-P-plots} and the related discussions in the
Appendix~\ref{sec:bi-p}. Associated questions on the effect of the
uncertainty of $\theta_{23}$ on $\delta_{\rm CP}$ determination in the
precision era has been addressed in \cite{Minakata:2013eoa}.

\subsection{Total of 10 running years  ($10^{22}~{\rm POT}$)}
\label{10years-T2K}

\begin{figure}[htbp]
\begin{center}
\vspace{2mm}
\includegraphics[width=0.31\textwidth]{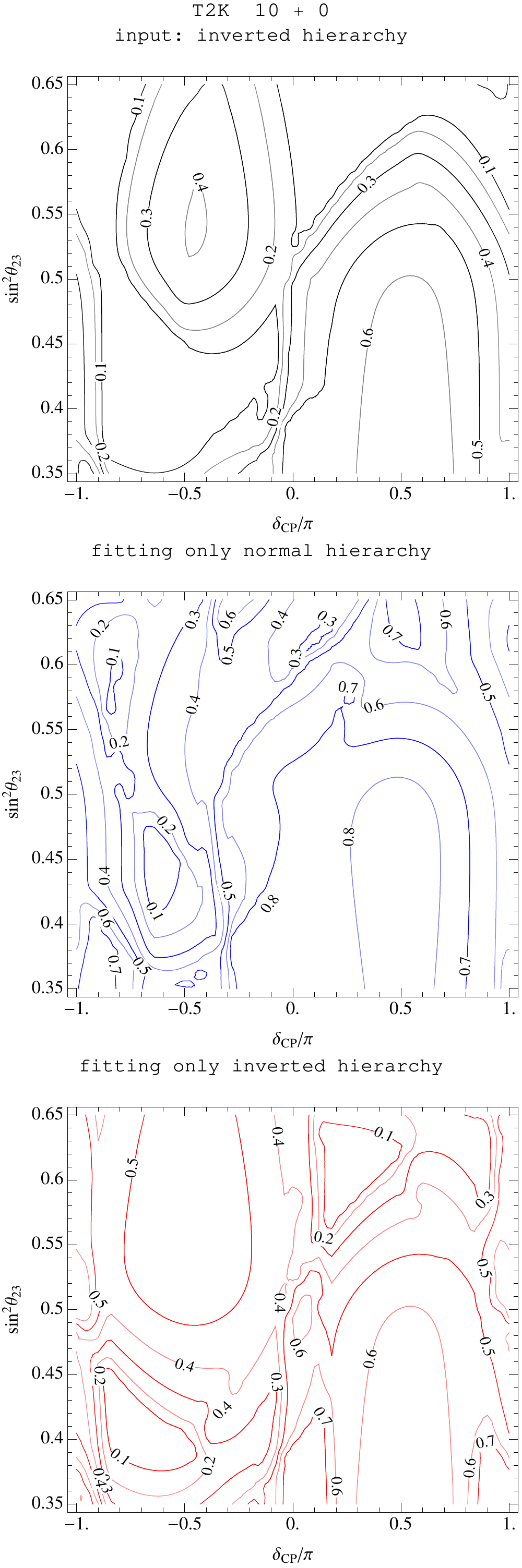}
\includegraphics[width=0.31\textwidth]{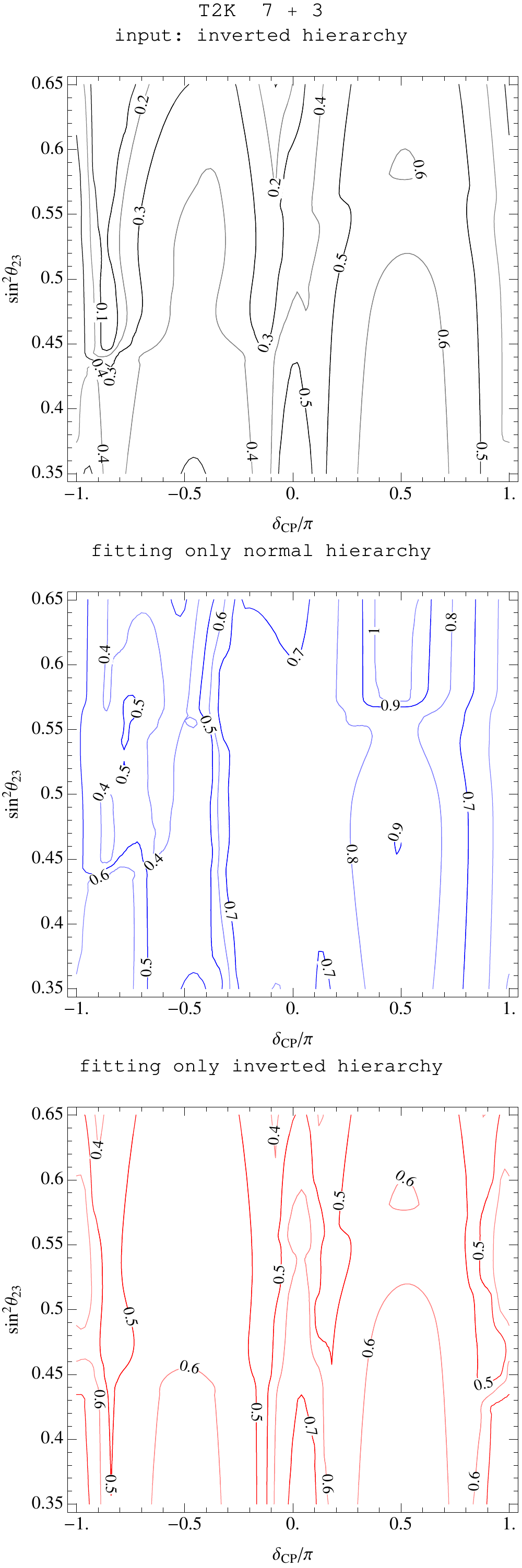}
\includegraphics[width=0.31\textwidth]{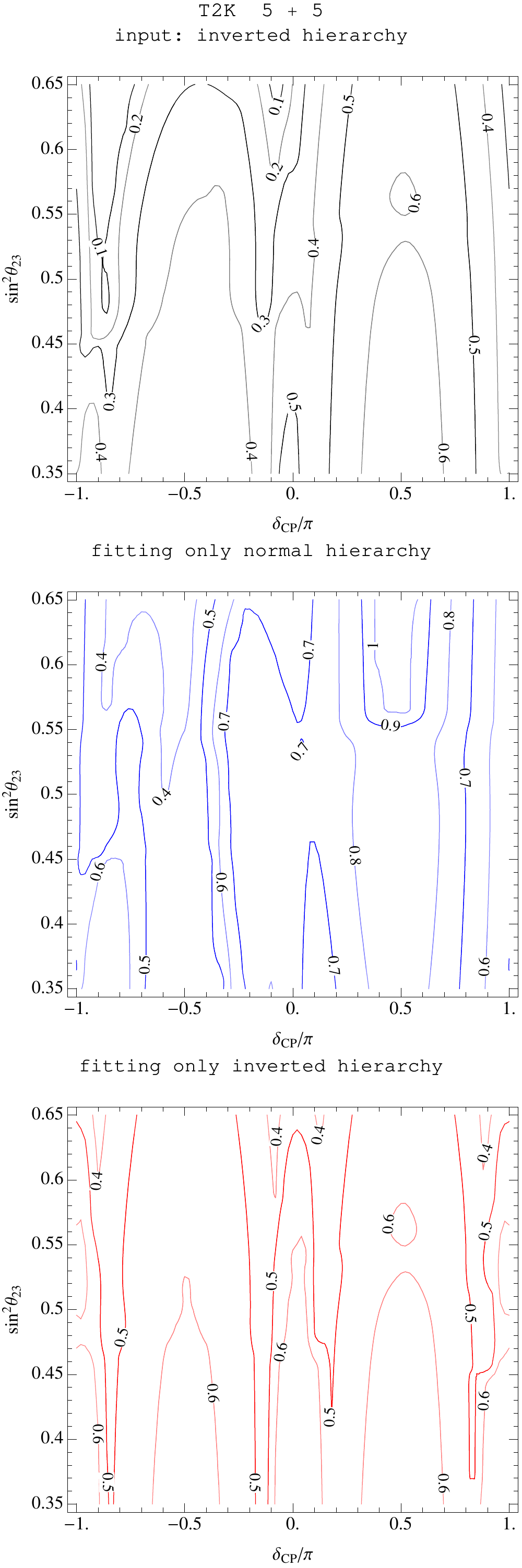}
\end{center}
\vspace{-7mm}
\caption{CP exclusion fraction isolines plotted on the
  $\delta_{\text{CP}} - \sin^2 \theta_{23}$ plane at 90 \% CL, for T2K
  running in $\nu$+$\bar \nu$ mode for $10+0$ (left), $7+3$ (center)
  and $5+5$ (right) years. The input mass hierarchy is the inverted
  one.  The top panels are for a fit marginalizing over the
  hierarchies, while the middle (bottom) panels are for a fit imposing
  the normal (inverted) hierarchy.  }
\label{T2K-10-IH}
\end{figure}

In Fig.~\ref{T2K-10-IH} we present similar contours of equal CP
exclusion fraction for a total of 10 running years with $\nu
+\bar{\nu}$ beam time sharing of $10+0$, $7+3$, and $5+5$ years
(panels from left to right), assuming the nominal design
luminosity. The results for $3+7$ running years (not shown) are
similar to the latter two cases, which represent the best
sensitivities among the studied cases of a total of 10 running
years. The results presented in the top panels were obtained by
marginalizing over the mass hierarchies (black contours). The
middle and bottom panels are for cases of a fit assuming the normal
(blue contours) and the inverted (red contours) mass
hierarchies, respectively.  In Fig.~\ref{T2K-10-IH}, only the case for
inverted mass hierarchy as input is shown.

The main features of the CP exclusion fraction contours for the normal
mass hierarchy as input may be obtained, in the zeroth order
approximation, by doing the re-parameterization $\delta_{\text{CP}}
\rightarrow \pi-\delta_{\text{CP}}$ in Fig.~\ref{T2K-10-IH}. This
approximation is valid because of the small matter effect in the T2K
setting. The particular case of T2K $5+5$ running years with the
normal hierarchy as input is shown in the next section, see
Fig.~\ref{NOVA-10}.

It should be emphasized first that as in the case of 5 years of data
taking, the inclusion of antineutrino running time significantly
improves the sensitivity to CP phase. Some of the distinctive features
of running T2K for 10 years, shown in Fig.~\ref{T2K-10-IH}, compared
to the results in 5 years running shown in Fig.~\ref{T2K-5}, are:

\begin{itemize}
\item 
With marginalization over the mass hierarchies (top panels) the null
sensitivity regions become significantly smaller, in
particular, if we compare the last two top panels of each figure.

\item 
The $7+3$ and $5+5$ years running results, when fitted assuming the
inverted mass hierarchy (the correct one in this case), can exclude
50\% (or higher) values of $\delta_{\rm CP}$ in almost the entire
$\delta_{\text{CP}} -\sin^2 \theta_{23}$ plane allowed by the current
oscillation data.  This can be seen in the bottom center and right
panels.

\item 
The $7+3$ and $5+5$ years running results, when fitted using the
  normal mass hierarchy, can exclude a fraction of  $\delta_{\rm CP}$
  values up to 80\%-90\% for $\delta_{\rm CP}>0$. The higher exclusion
  power is due to the assumption of the wrong mass hierarchy. But for
  $\delta_{\rm CP}<0$, specially when $\theta_{23}$ is in the second
  octant, the exclusion fraction tends to be much less than the one
  for the right hierarchy.
\end{itemize}
\noindent

What is the meaning of doing a fit assuming the wrong mass hierarchy?
If the hierarchy is known with high confidence level, of course, there
is no physics sense of attempting a fit assuming the wrong mass
hierarchy. The real question is: What does it mean at the time in
which the mass hierarchy is not established? We argue that it is an
alternative and useful way of probing the mass hierarchy sensitivity
in terms of the CP exclusion fraction. Since this point will become
clearer in the discussion of NO$\nu$A results we will come back to it
in the next section.

\section{Sensitivity to CP phase expected by NO$\nu$A and by Its Combination with T2K}
\label{CP-NOVA}

\subsection{10  running years: NO$\nu$A ($6\times 10^{21}~{\rm POT}$)}

\begin{figure}[htbp]
\begin{center}
\vspace{2mm}
\includegraphics[width=0.31\textwidth]{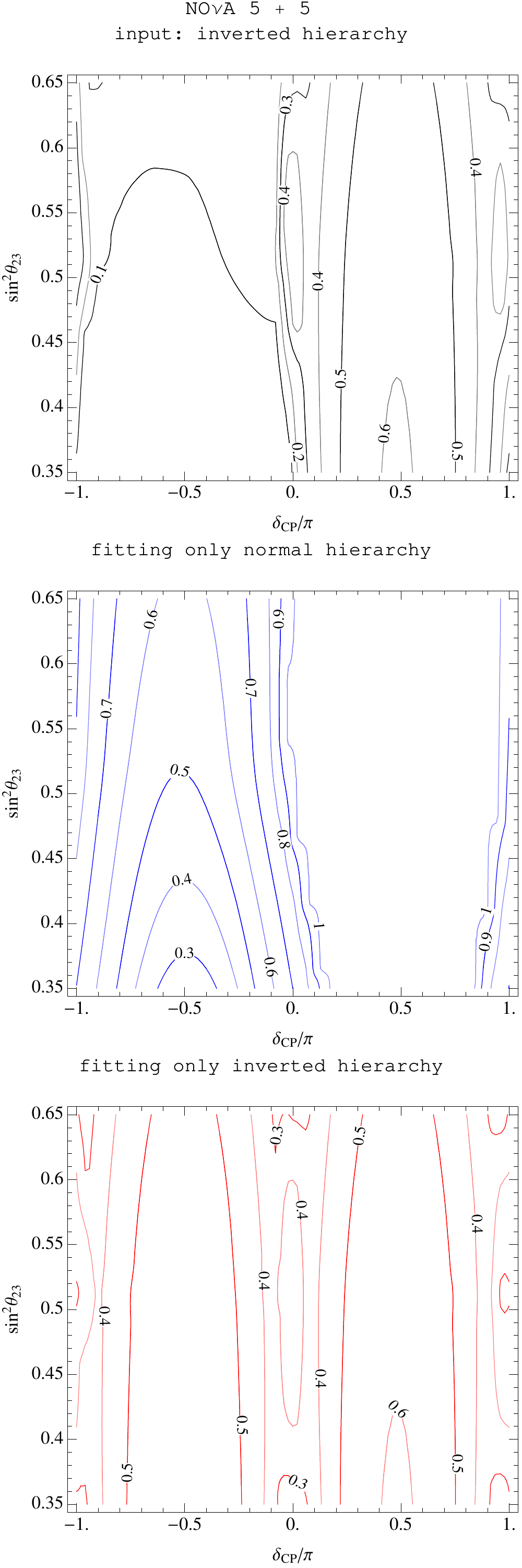}
\includegraphics[width=0.31\textwidth]{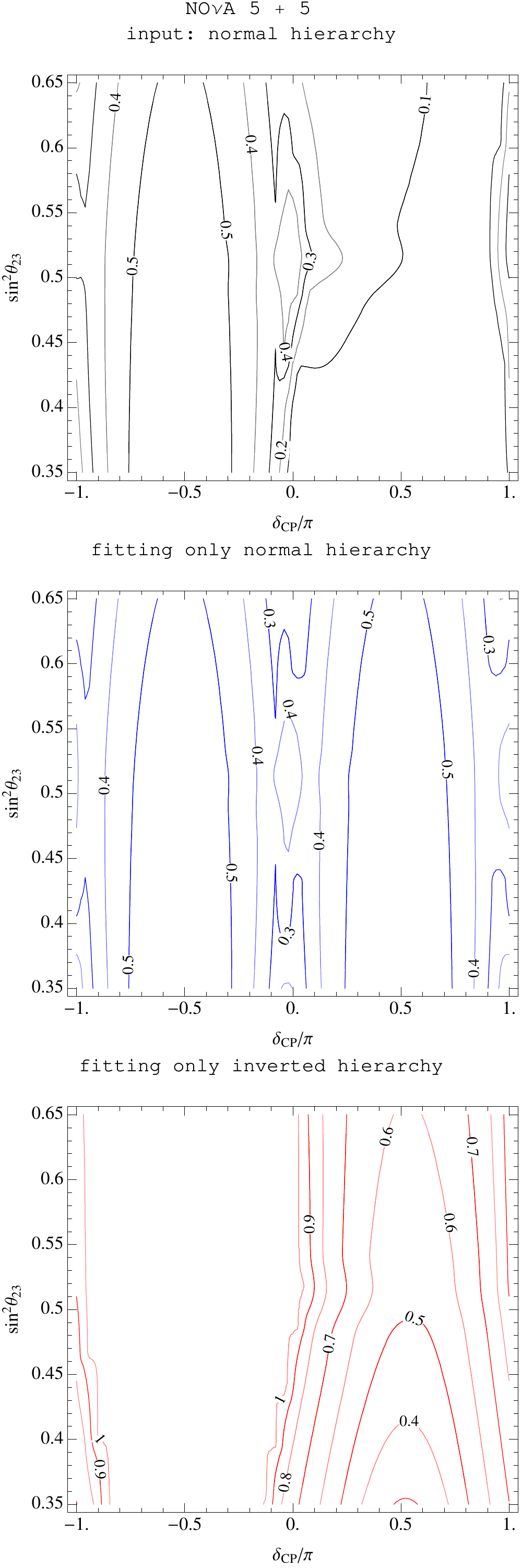}
\includegraphics[width=0.31\textwidth]{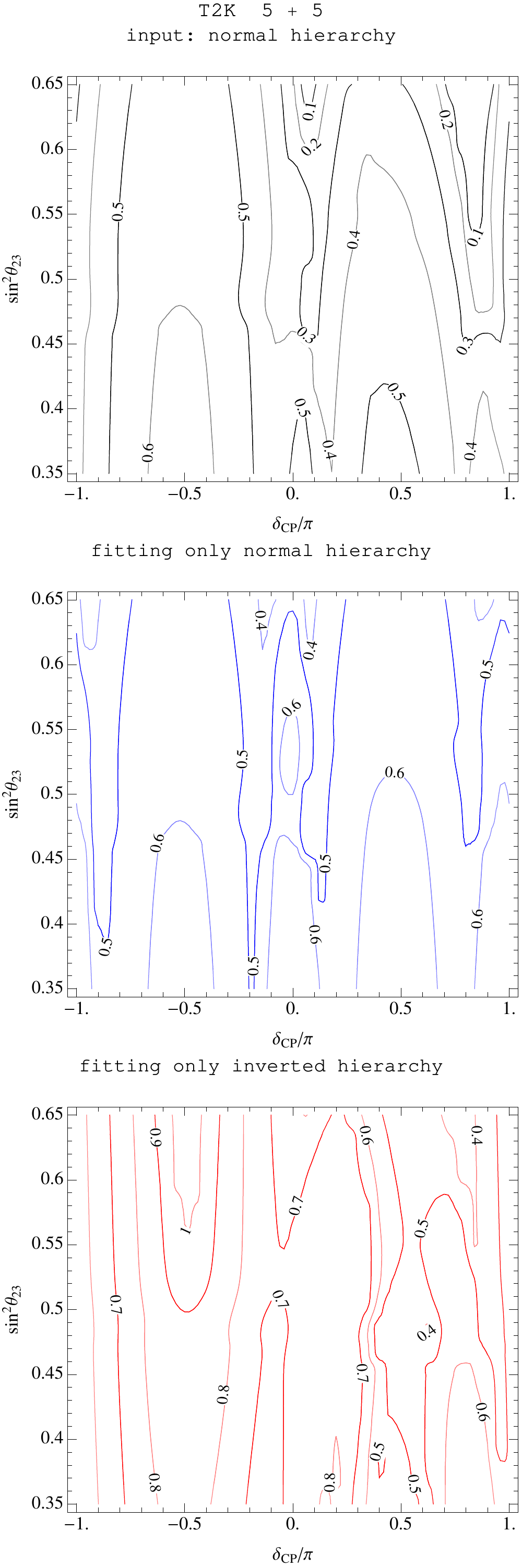}
\end{center}
\vspace{-7mm}
\caption{CP exclusion fraction isolines plotted on the
  $\delta_{\text{CP}} - \sin^2 \theta_{23}$ plane at 90 \% CL, for
  NO$\nu$A and T2K running in $\nu$+$\bar \nu$ mode for $5+5$
  years. The left and center panels are for NO$\nu$A with inverted and
  normal mass hierarchy as input.  The right panels are for T2K with
  normal mass hierarchy as input.  The top panels are for a fit
  marginalizing over the hierarchies, while the middle (bottom) panels
  are for a fit imposing the normal (inverted) hierarchy.  }
\label{NOVA-10}
\end{figure}

In Fig.~\ref{NOVA-10} the contours of equal CP exclusion fraction are
plotted for a total of 10 running years of the NO$\nu$A experiment
with $\nu + \bar{\nu}$ beam time sharing of $5+5$ years. The left and
middle panels are for the case of inverted and normal mass
hierarchies, respectively. The results for $7+3$ years running (not
shown) are similar to the ones in Fig.~\ref{NOVA-10}.  To make a
comparison with T2K sensitivity to CP phase easier we place on the
right panels of Fig.~\ref{NOVA-10} the contours of equal CP exclusion
fraction obtained by T2K $5+5$ years running in the case of normal
input mass hierarchy. (For similar contours with the inverted mass
hierarchy see Fig.~\ref{T2K-10-IH}.) As in Fig.~\ref{T2K-10-IH} the
upper panels are for cases marginalized over the mass hierarchies
(black contours). The middle and bottom panels are for cases of
a fit assuming the normal (blue contours) and the inverted
(red contours) mass hierarchies, respectively.

We notice the following two significant features of NO$\nu$A's CP
sensitivity in comparison to that of T2K:\footnote{
One should keep in mind that we are comparing between the
sensitivities expected by these two experiments by taking the
particular official values of neutrino fluxes, $10^{21}$ and
$6 \times 10^{20}$ POT a year for T2K and NO$\nu$A, respectively.  }

\begin{itemize}
\item 
The sensitivity of NO$\nu$A to CP phase is worse than that of T2K when
marginalized over the mass hierarchies (top panels), almost losing the
sensitivity in the negative (positive) half plane of $\delta_{\rm CP}$
for the input inverted (normal) mass hierarchy.

\item 
Similarly, T2K is slightly better than NO$\nu$A in the CP sensitivity
assuming the right mass hierarchy (middle panels of the second and
third columns), having 60\% contours of CP exclusion in both half
planes of $\delta_{\text{CP}}$. On the other hand, in the wrong mass
hierarchy fit the NO$\nu$A CP sensitivity is overwhelming, making
almost a complete exclusion at 90\% CL of one of the half planes
possible.

\end{itemize}
\noindent
Let us understand these characteristics. It appears that the
relatively low NO$\nu$A CP sensitivity compared to that of T2K comes
partly from the relatively low statistics. Although the number of
events depends on the input parameters, it appears that in general T2K
is able to accumulate 20--30\% more statistics than NO$\nu$A.  In
addition to this, as discussed in Appendix~\ref{sec:bi-p}, the fact
that the major axis of the CP ellipse for NO$\nu$A is shorter than
that for T2K (see Fig.~\ref{Bi-P-plots}) makes the CP sensitivity of
NO$\nu$A worse than that of T2K, even for similar statistics.

On the other hand, the powerfulness of excluding almost half the space
(positive $\delta_{\text{CP}}$ region for the inverted, and negative
$\delta_{\text{CP}}$ region for the normal mass hierarchies) in the
wrong hierarchy fit is due to the larger matter effect thanks to the
longer baseline of NO$\nu$A.  Using this property the CP exclusion
fraction may be used as a powerful indicator of the mass hierarchy
though in a particular region of $\delta_{\text{CP}}$. Therefore, it
appears to us that these two experiments complement each other quite
nicely.

\begin{figure}[htbp]
\begin{center}
\vspace{2mm}
\includegraphics[width=0.31\textwidth]{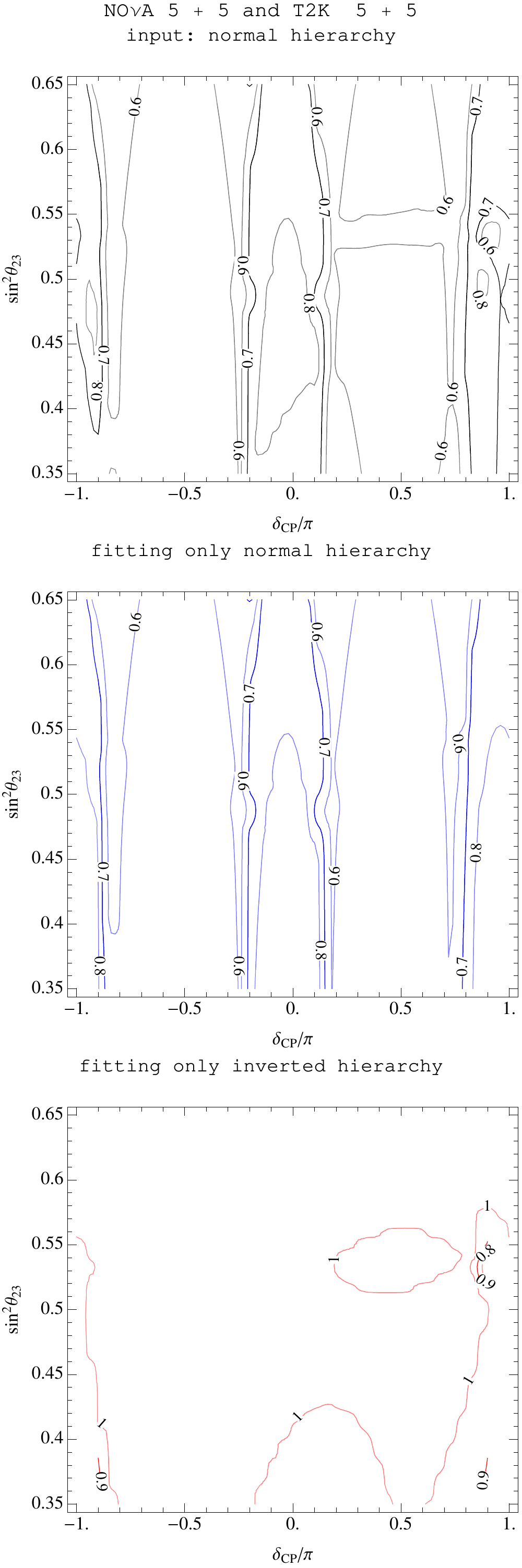}
\includegraphics[width=0.31\textwidth]{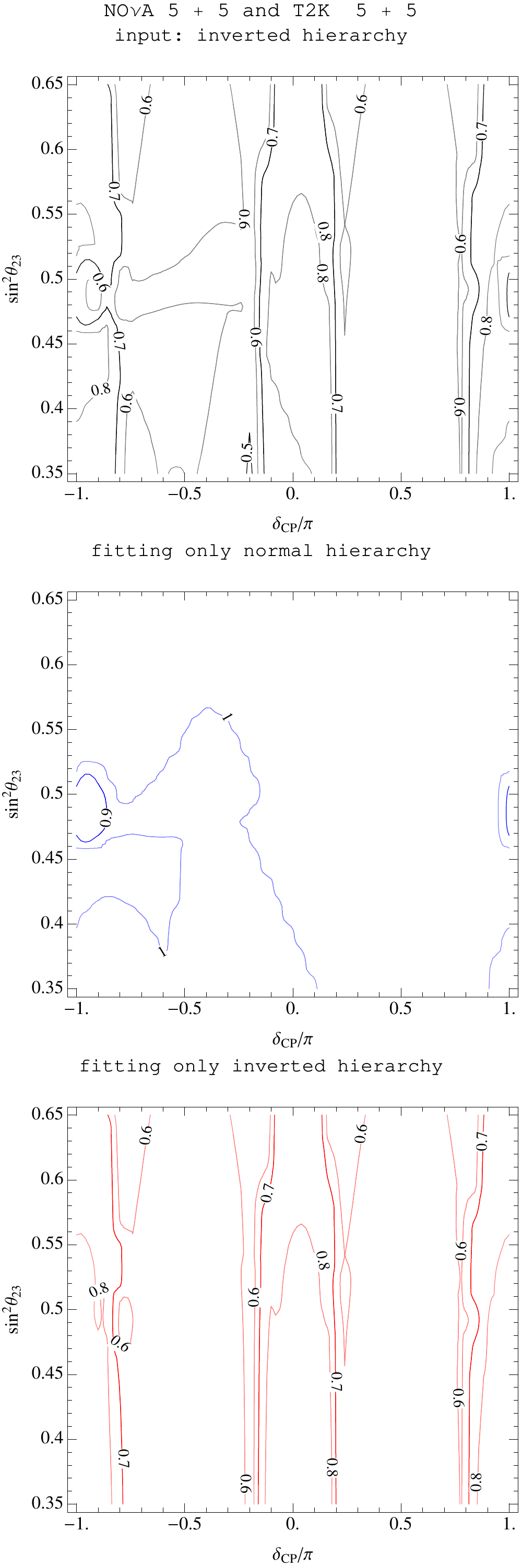}
\includegraphics[width=0.31\textwidth]{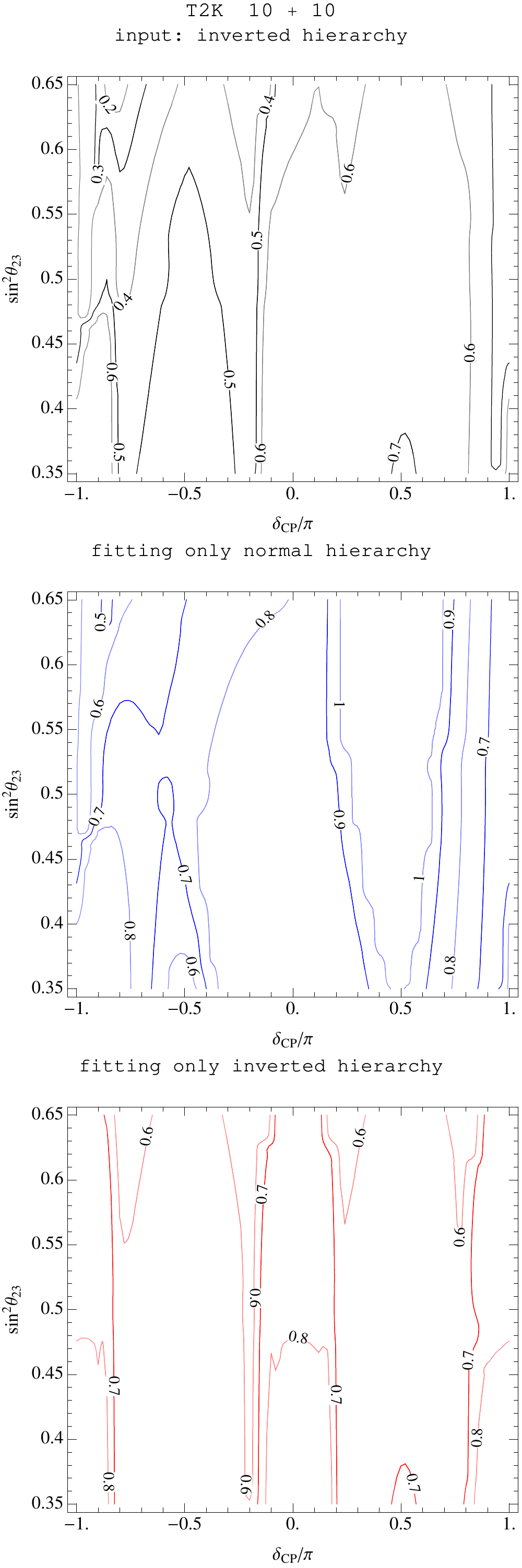}
\end{center}
\vspace{-6mm}
\caption{CP exclusion fraction isolines plotted on the
  $\delta_{\text{CP}} - \sin^2 \theta_{23}$ plane at 90 \% CL, for
  NO$\nu$A and T2K running in $\nu$+$\bar \nu$ mode for $5+5$ years
  (each) combined as well as T2K running for $10+10$ years. The left
  and center panels are for the combination using the inverted and
  normal mass hierarchy, respectively, as input.  The right panels are
  for T2K with inverted mass hierarchy as input.  The top panels are
  for a fit marginalizing over the mass hierarchies, while the middle
  (bottom) panels are for a fit imposing the normal (inverted) mass
  hierarchy.}
\label{T2K-NOVA}
\end{figure}

\subsection{Combination of NO$\nu$A with T2K and the synergy}

One of the most intriguing questions would be how high is the sensitivity to the
CP phase when T2K and NO$\nu$A are combined, and to what extent a synergy can be
expected. To answer these questions, we present in Fig.~\ref{T2K-NOVA} the
contours of CP exclusion fraction obtained by combining $5+5$ years running of
T2K and NO$\nu$A (a total of 10 years each) for the input normal (left panels)
and inverted (middle panels) mass hierarchies. To extract the effect of the
synergy we place in the right panel of Fig.~\ref{T2K-NOVA} the contours obtained
by a hypothetical $10+10$ years running of T2K (a total 20 years). Although we
do not consider it a realistic option, we show it for the sake of revealing the
synergy.

The distinctive features of Fig.~\ref{T2K-NOVA} are as follows:

\begin{itemize}

\item 
One of the most significant features in Fig.~\ref{T2K-NOVA} is that the wrong
mass hierarchy is excluded at 90\% CL in almost the entire allowed region in
$\delta_{\text{CP}} - \sin^2\theta_{23}$ space. 
In the settings discussed in this paper it can only be
achieved by combining T2K and NO$\nu$A.

\item 
It is quite remarkable in the left  and center
  set of panels that in the entire $\delta_{\text{CP}} -
\sin^2\theta_{23}$ space is covered by 60\% or higher exclusion
fraction region, even with marginalization over the mass hierarchies,
or in the right mass hierarchy fit.

\item 
For the case of T2K and NO$\nu$A combined or the T2K only 
but for the known mass hierarchy,  
the region of the highest sensitivity tends to exist 
at $\delta_\text{CP} \sim 0$ or $\pm \pi$, which is different
from the cases of lower statistics where the highest sensitivity 
likely to occur at $\delta_\text{CP} \sim \pm \pi/2$. 
A qualitative explanation of this feature based on 
the bi-probability plots is found in Appendix~\ref{sec:bi-p}. 

\item 
Another salient feature in Fig.~\ref{T2K-NOVA} is that the effect of
the synergy is evident when T2K and NO$\nu$A combination (each for a
total of 10 years, left and center panels) is compared to T2K running
for 20 years (right panels). This is so, in particular, in the case
with  marginalization over the mass hierarchies, or for the wrong
mass hierarchy fit.
\end{itemize}

\section{The interplay between $\dcp$ and $\theta_{23}$ octant for the
  experimental strategy}
\label{result-23}

Until now, we have focused on the sensitivity to CP phase and discussed some
strategy to optimize it. Actually, T2K and \nova can endeavor to measure another
very important unknown: the octant of $\theta_{23}$. That said, we raise the
straightforward question ``How the strategies for determining $\dcp$ and the
$\theta_{23}$ octant are related?''\footnote{
Note, however, that solving the $\theta_{23}$ octant degeneracy is not the whole story, as stressed in \cite{Minakata:2013eoa}. } 
See Ref.~\cite{Agarwalla:2013ju,Chatterjee:2013qus} which also discussed the octant determination by combining T2K and \nova. 

To answer this question, let us first recollect some relevant features of the
$\theta_{23}$ octant measurement.  Due to high statistics of the disappearance
channels $\nu_\mu\to\nu_\mu$ and $\bar\nu_\mu\to\bar\nu_\mu$, $\sin^2
2\theta_{23}$ can be measured with high precision, but they are insensitive to
the $\theta_{23}$ octant. On the other hand, because of its relatively low
statistics, the appearance channels $\nu_\mu\to\nu_e$ and
$\bar\nu_\mu\to\bar\nu_e$ have a capability of breaking the octant degeneracy
\emph{only if} the determinations of $\sin^2 2\theta_{13}$ and $\sin^2
2\theta_{23}$ are precise enough.  For concreteness, let us focus on T2K.  After
10 years of running, we expect that the determination of $\theta_{23}$ by the
disappearance channels is dominated by  systematic errors. Hence its
sensitivity to $\sin^2 2\theta_{23}$ would be approximately independent of the
running configuration.

Now, we discuss how to proceed with the $\nu_{e}$ and $\bar{\nu}_{e}$
appearance channels. If T2K runs solely in the neutrino mode, the
octant degeneracy becomes virtually unsolvable even if we take into
account energy spectrum as we will see below. First, let us consider
only the total rates. From Fig.~\ref{Bi-P-plots}, we can see that by
only using the neutrino mode, even if we know the true mass hierarchy
and the precise value of the oscillation probability, $P(\nu_\mu \to
\nu_e)$, unless we know rather well the value of $\delta_{\text{CP}}$,
$\theta_{23}$ different octants can be confused. This is in general
true apart from the case where $\theta_{23}$ lies in the 1st (2nd)
octant and $\delta_{\text{CP}}$ is close to $\pi/2$ ($-\pi/2$).

\begin{figure}[t]
\vglue -2cm
\begin{center}
\hglue 0.2cm
\includegraphics[width=0.8\textwidth]{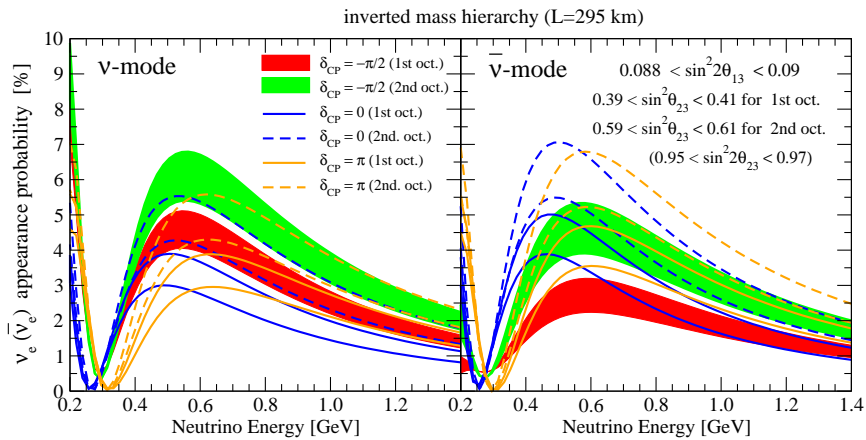}
\end{center}
\vglue -1.5cm 
\caption{Appearance probabilities $P(\nu_\mu\to\nu_e)$ for neutrino
  (left panel) and $P(\bar{\nu}_\mu\to\bar{\nu}_e)$ for antineutrino
  (right panel) as a function of the neutrino energy, for
  $\delta_{\text{CP}} = 0, \pm \pi/2 $ for the case where $0.95 <
  \sin^2 2 \theta_{23} < 0.97$.
}
\label{fig:prob-energy}
\vspace{-1mm}
\end{figure}

We can try to see if the energy spectrum information will help in resolving this
degeneracy.  Let us look at Fig.~\ref{fig:prob-energy} which shows the
appearance probabilities $P(\nu_\mu\to\nu_e)$ for neutrino (left panel) and
$P(\bar{\nu}_\mu\to\bar{\nu}_e)$ for antineutrino (right panel) as a function of
the neutrino energy for the case where $0.95 < \sin^ 2 2\theta_{23} < 0.97$ for
various different values of $\delta_{\text{CP}}$.  We can see from the left
panel of Fig.~\ref{fig:prob-energy} that the two cases of $\delta_{\text{CP}} =
-\pi/2$ with $\theta_{23}$ in the 1st octant (filled band by red color) and
$\delta_{\text{CP}} = 0$ with $\theta_{23}$ in the 2nd octant (the band
delimited by the dashed blue curves) are easily confused even if we take into
account the energy spectrum.
However, these two cases give very different probabilities in the antineutrino
modes, as we can see from the right panel of Fig.~\ref{fig:prob-energy}, which
means that combining with antineutrino data will certainly help in resolving the
octant degeneracy.  The importance of the antineutrino run in resolving the
octant degeneracy was inherent in the analysis in Ref.~\cite{Hiraide:2006vh},
and some of the related points are discussed recently in
Ref.~\cite{Agarwalla:2013ju,Chatterjee:2013qus}.

\begin{figure}[t!]
\begin{center}
\vspace{1mm}
\includegraphics[width=0.31\textwidth]{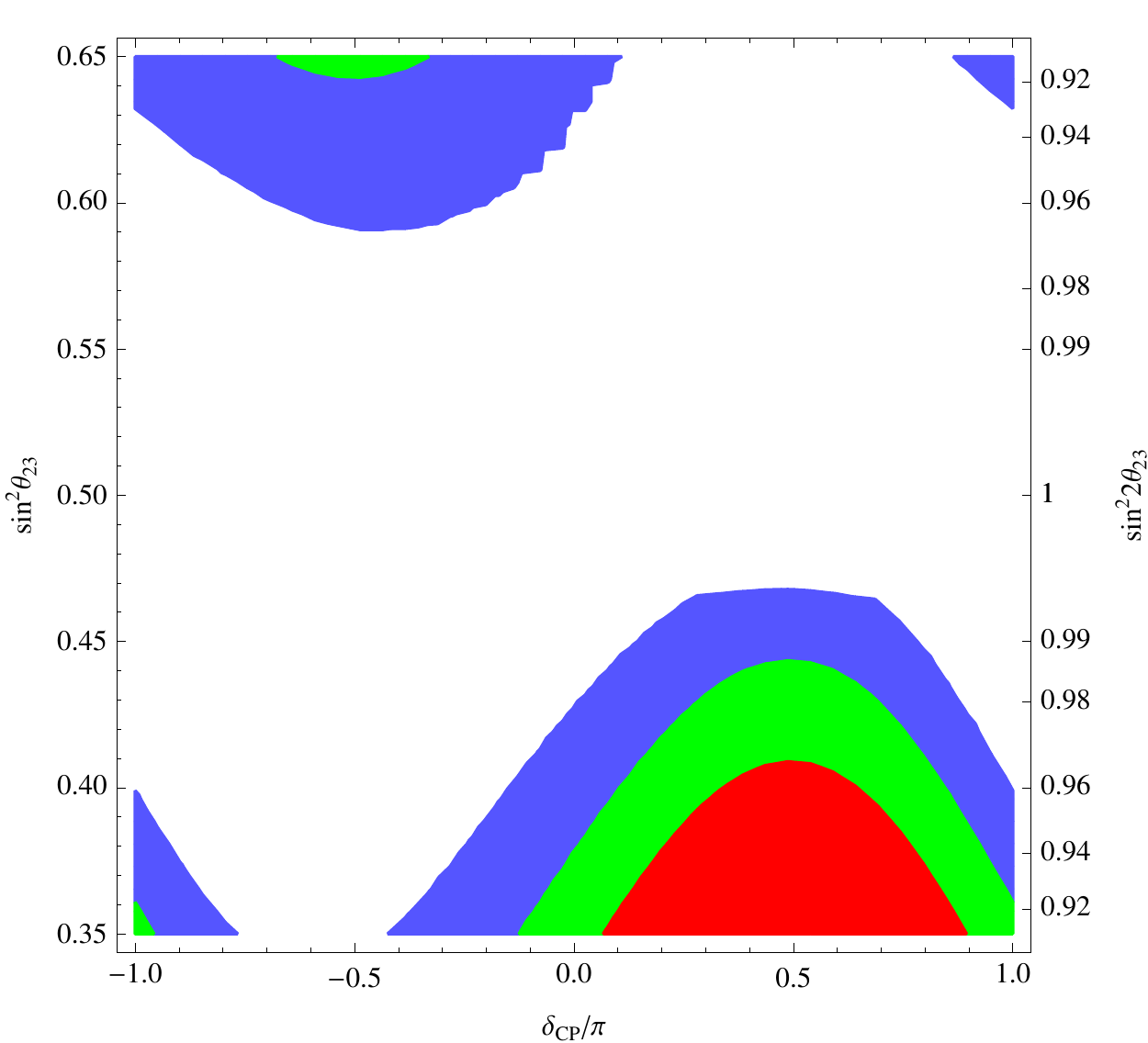}
\includegraphics[width=0.31\textwidth]{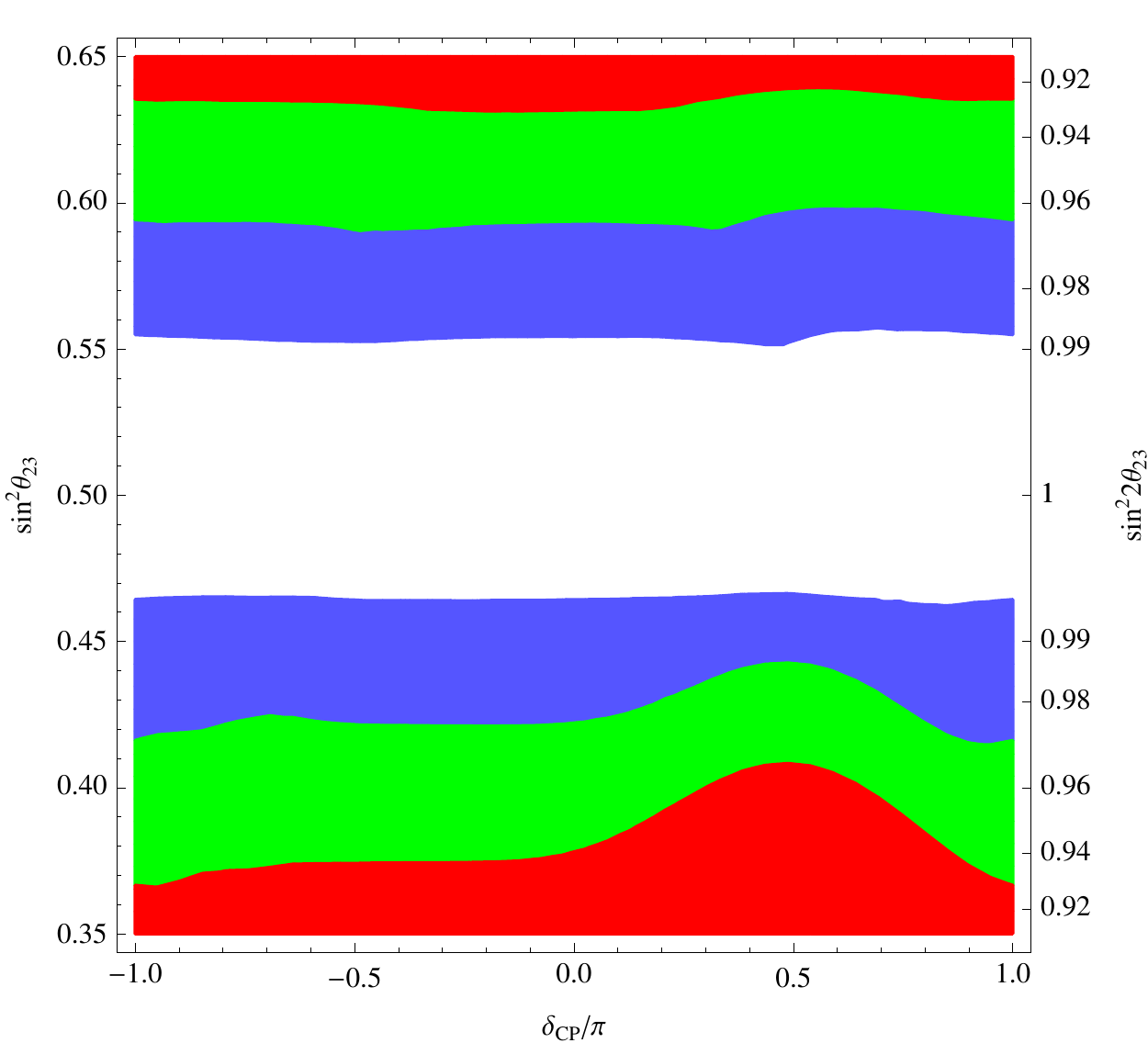}
\includegraphics[width=0.31\textwidth]{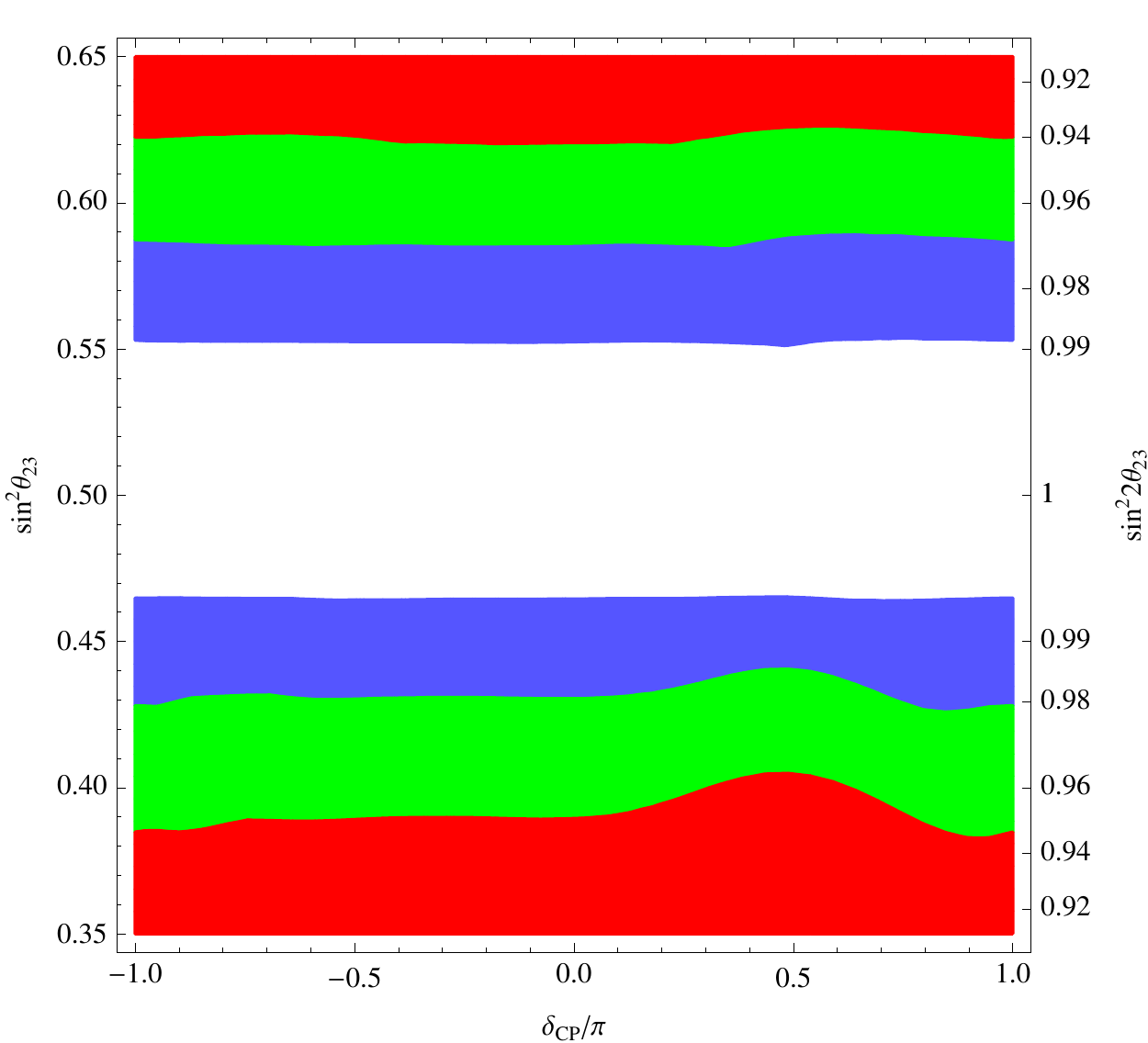}
\includegraphics[width=0.31\textwidth]{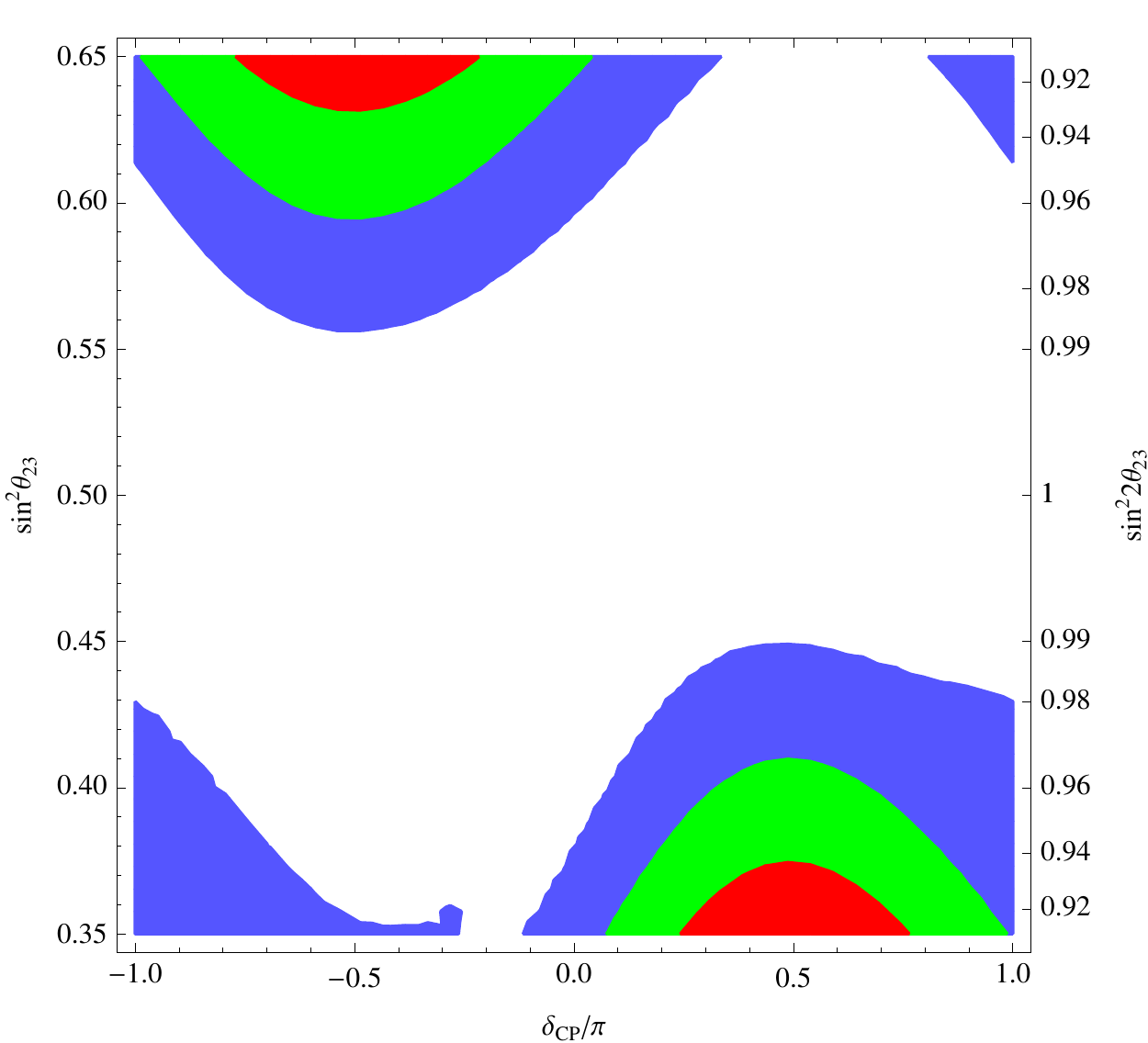}
\includegraphics[width=0.31\textwidth]{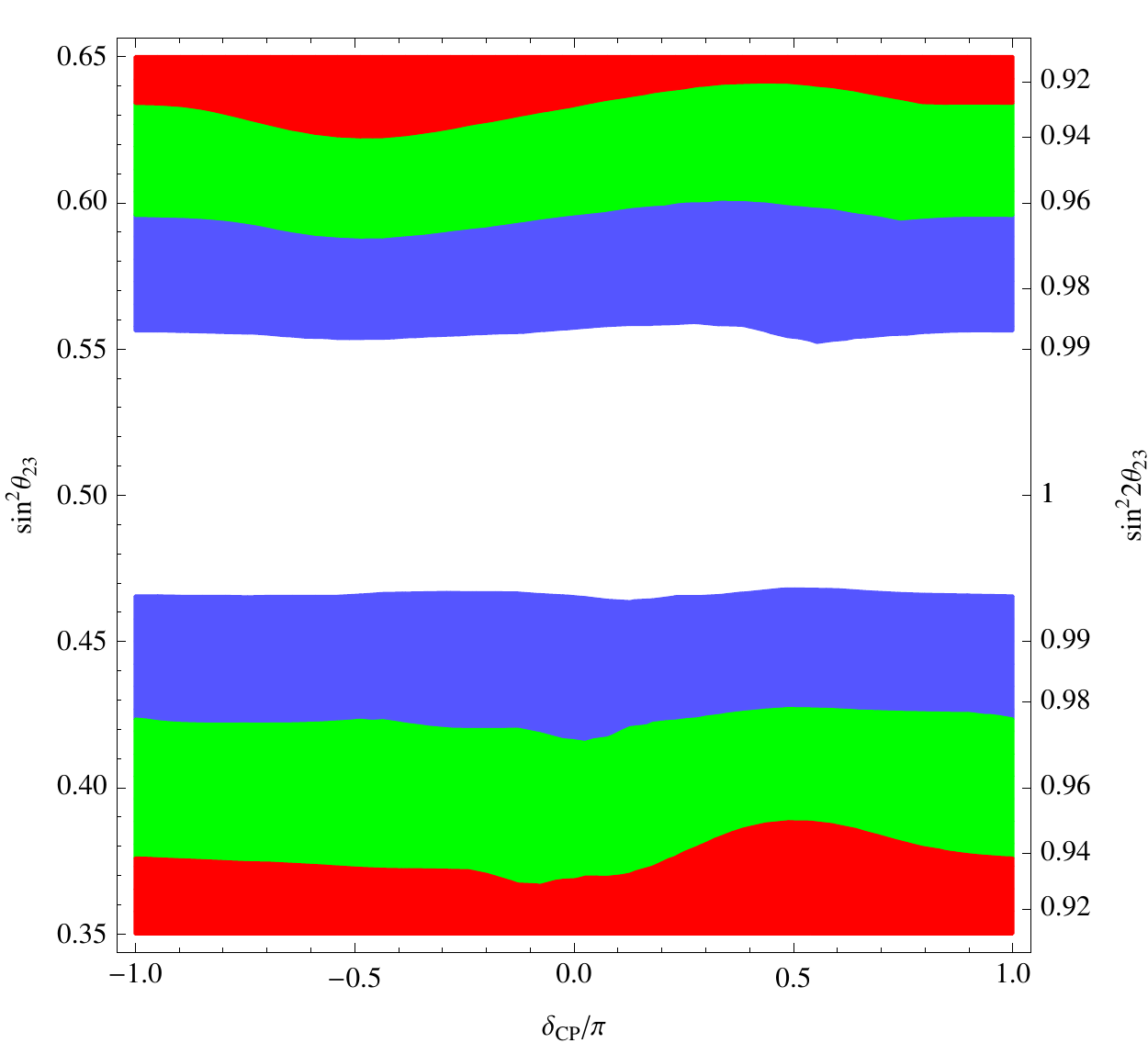}
\includegraphics[width=0.31\textwidth]{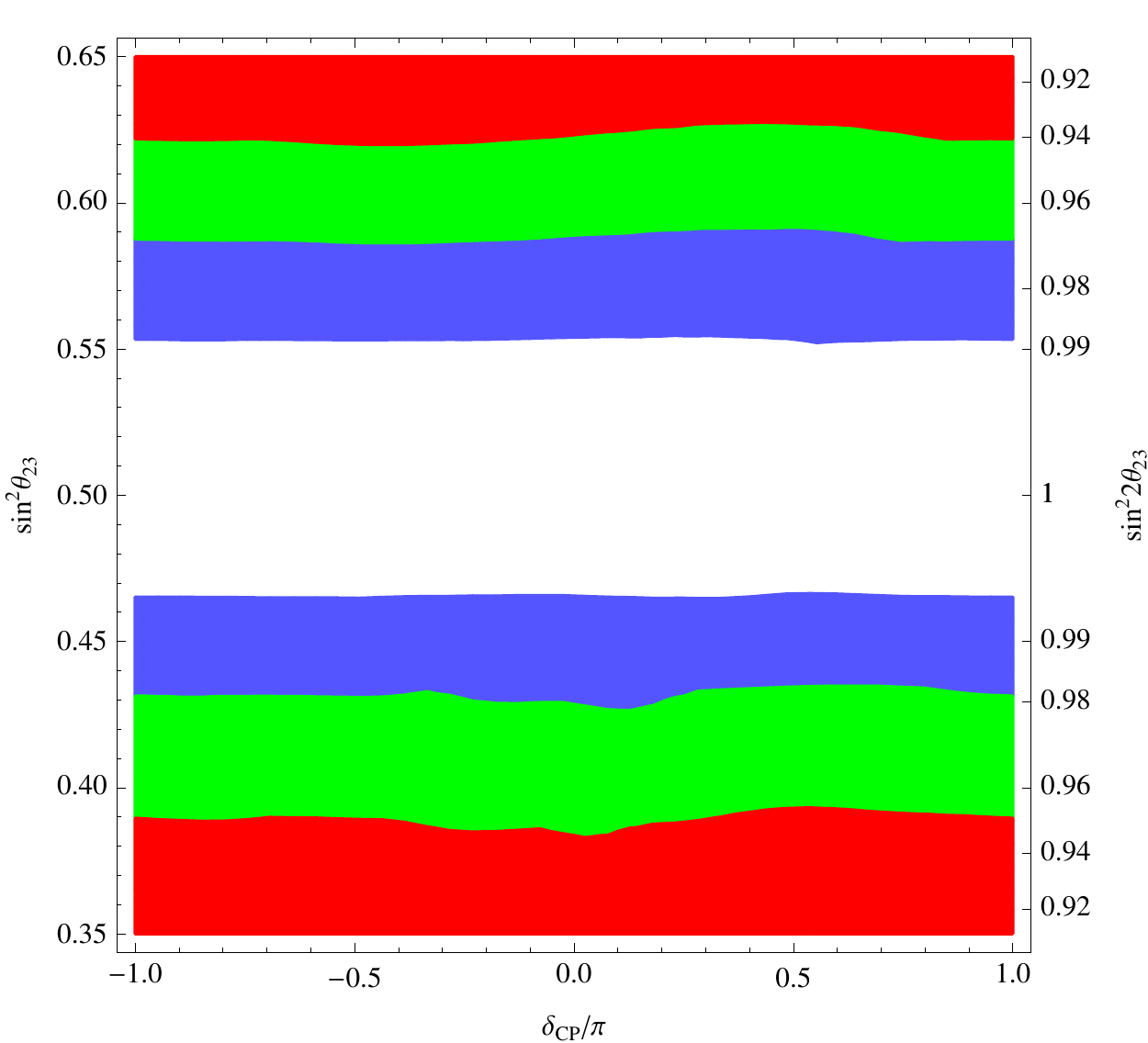}
\end{center}
\vspace{-6mm}
\caption{Regions in which the $\theta_{23}$ octant
    degeneracy is resolved are plotted on the $\delta - \sin^2
    \theta_{23}$ plane for $10+0$ (left panel), $7+3$ (middle panel),
    $5+5$ (right panel) years of $\nu$ + $\bar{\nu}$ running of T2K. 
  In the upper and lower panels the cases of inverted and normal mass
  hierarchies, respectively, are shown.
}
\label{T2K-23-IH}
\end{figure}

When the antineutrino running is incorporated in T2K, the comparison
between the event rates as well as the energy spectra of the $\nu$ +
$\bar{\nu}$ modes challenges the degeneracy toward its resolution in a
more robust way.  In order to understand to what extent the mechanism
works, we present in Fig.~\ref{T2K-23-IH} the regions of resolution of
the octant degeneracy in $\delta_{\text{CP}} - \sin^2 \theta_{23}$
space, calculated by imposing a Gaussian uncertainty in $\sin^2
2\theta_{23}$ of $0.02$ at 68\% CL.  The regions colored in blue,
green and red represent the region on the plane of the true values of
$\dcp$ and $\sin^2 \theta_{23}$ in which the octant of $\theta_{23}$
can be distinguished at 1$\sigma$, 2$\sigma$, and 3$\sigma$ CL,
respectively. Around maximal $\theta_{23}$, no identification of the
$\theta_{23}$ octant is possible even at 1$\sigma$ CL. In the panels
from left to right are shown $10+0$, $7+3$, and $5+5$ years
running. It is also worth mentioning that for $3+7$ years running
almost the same sensitivity as for running $7+3$ or $5+5$ years is
obtained.

As can be seen, the inclusion of the antineutrino run significantly
improves the sensitivity to the octant determination of $\theta_{23}$
because it can break the octant degeneracy as discussed above. We also
notice that once the fraction of time allocated for antineutrino
running reaches 30\% or so of the total running time the sensitivity
to the octant of $\theta_{23}$ is remarkably stable against variations
of this fraction.  As we saw, the impact of the $\nu$ + $\bar{\nu}$
beam time sharing on the octant determination is much less important
than on the CP sensitivity, at least for the experiments we are
interested in.  It seems to us that the optimal proportion of
antineutrino to neutrino could be mainly dictated by the sensitivity
for the CP phase without loosing essentially that for the octant
$\theta_{23}$ determination.

\section{Conclusion}
\label{conclusion}

In the near future, 5 to 10 years from now, we do not expect to be
able to measure the lepton CP phase, since we will not yet dispose of
neutrino experiments designed to discover CP violation due to non-zero
$\sin\delta_{\rm CP}$. However, the accelerator based neutrino
oscillation experiments, T2K and NO$\nu$A, after the precise
measurement of $\sin^2 \theta_{13}$ by the reactor experiments, will
have some sensitivity to $\delta_{\rm CP}$. This sensitivity will
depend on the true values of $\delta_{\rm CP}$, $\sin^2 \theta_{23}$,
the neutrino mass hierarchy as well as the amount of data taking in
neutrino and antineutrino modes.

To study the maximal sensitivity to $\delta_{\rm CP}$ attainable by a
single or a set of experiments we employed the CP exclusion fraction,
which quantifies the range of $\delta_{\rm CP}$ that can be excluded,
at a certain confidence level (we adopted 90\% in this paper), by a
set of experimental observables. We expect that the CP exclusion
fraction is particularly useful to examine the potential of exploring
CP phase possessed by the near future experiments which may be the
unique sources of information on the CP phase in an era without
dedicated apparatus designed for the discovery of CP violation.

By using the CP exclusion fraction we have analyzed the CP sensitivity
of T2K and NO$\nu$A experiments. We have shown that it is important to
run T2K in the antineutrino mode in order to significantly enhance the
CP sensitivity of this experiment. The optimal situation seems to be
to share the time equally between neutrino and antineutrino beams. If
one could run T2K for 10 years one would be able to exclude 50\% or
more of the $\delta_{\rm CP}$ values in almost the entire half plane
of $\delta_{\rm CP}>0$ ($\delta_{\rm CP}<0$) for the inverted (normal)
mass hierarchy and $\sin^2 \theta_{23} \in [0.35, 0.65]$, even not
knowing the neutrino mass hierarchy.  If the neutrino mass hierarchy
is known by that time, one could exclude 50\% (or more) of the
$\delta_{\rm CP}$ values on almost the entire $\delta_{\rm CP}-\sin^2
\theta_{23}$ plane.

We have shown that NO$\nu$A is less powerful than T2K for the CP
sensitivity as measured with the CP exclusion fraction. An intuitive
understanding of this feature is offered by using the bi-probability
plot and by noting the difference in statistics in both
experiments. By combining both experiments, however, one could exclude
60\% or more of $\delta_{\rm CP}$ values in the $\delta_{\rm CP} -
\sin^2 \theta_{23}$ plane. It should be noticed that the synergy
between these two experiments is quite visible, allowing the exclusion of
the wrong mass hierarchy at 90\% CL in almost the entire $\delta_{\rm
  CP}-\sin^2 \theta_{23}$ space.

We have also examined T2K sensitivity to the $\theta_{23}$ octant,
showing that adding antineutrino run also helps the experimental
sensitivity to $\sin^2 \theta_{23}$. The determination of this
parameter will further help constraining $\delta_{\rm CP}$, as it will
exclude part of the currently allowed region of $\delta_{\rm CP}$ and
$\sin^2 \theta_{23}$.

We emphasize that the 10\% uncertainty we adopt in our analyses, for
both experiments, may be a very conservative choice, in particular for
the analysis of 10 years running. This is because T2K already achieved
the uncertainty of $\simeq 10\%$ for running in neutrino mode, and it
is conceivable that this will be improved in the future. A caution is,
however, that so far little experimental information is accumulated in
the antineutrino mode.

The results of our analysis in this paper underlines the necessity of
dedicated experiments specially designed to access the lepton CP
violating phase $\delta_{\rm CP}$. Examples for such apparatus include
Hyper-Kamiokande or LBNE. Nonetheless, we emphasize the importance of
getting as much information as we can on $\delta_{\rm CP}$ before that
day of dedicated machines arrives. It will certainly help us to lay
the foundations for winning perhaps the long-term hardest job of
hunting the lepton CP phase, {\em the marathon in neutrino physics}.

\appendix
\section{Definition of CP exclusion fraction and its properties}
\label{fCPX}

\subsection{CP exclusion fraction; Definition}
\label{fCPX-def}

We follow the conventional $\chi^2$ method to calculate the likelihood, at a
given confidence level, of rejecting points in the parameter space $(\sin^2
\theta_{23}, \delta_{\rm CP})$ for a given input value of the parameters
$(\sin^2 \theta_{23}^{\rm in}, \delta_{\rm CP}^{\rm in})$.
Toward the goal, we compute the expected number of events $T_i$ in the
$i$-th energy bin as a function of the input parameters,
$T_i(\theta_{13}^{\rm in},\theta_{23}^{\rm in}, \delta_{\rm CP}^{\rm
  in}, h^{\rm in})$, where $h^{\rm in}$ is the input neutrino mass
hierarchy.  We also compute the expected number of events $F_i$ in the
$i$-th energy bin for a given set of fit and nuisance parameters
$\{\alpha\}$, $F_i(\theta_{13}^{\rm fit}, \theta_{23}^{\rm fit},
\delta_{\rm CP}^{\rm fit}, h^{\rm fit},\{\alpha\})$.  These numbers
include neutrino and antineutrino events, according to the assumed
exposure. With these we can build the likelihood function

\begin{eqnarray}
-2 \ln {\mathcal L}(\theta_{23}^{\rm in},\delta_{\rm CP}^{\rm in},h^{\rm in} ,\delta_{\rm CP}^{\rm fit})&=& \min_{\{\theta_{13}^{\rm fit}, \theta_{23}^{\rm fit}, h^{\rm fit},
\{\alpha\}\}} \biggl\{ 
\sum_{i=1}^{\rm nb} 2 \left( F_i-T_i + T_i \ln \frac{T_i}{F_i} \right) \nonumber 
+ \sum_j \left( \frac{\alpha_j}{\sigma_j} \right)^2 \\
&+& \left( \frac{\sin^22\theta_{13}^{\rm in}- 
\sin^22\theta_{13}^{\rm fit}}{\sigma_{13}} \right)^2
\biggr\} \, ,
\label{likelihood}
\end{eqnarray}
where we set 
$\sin^22\theta_{13}^\text{in} = 0.089$~\cite{An:2012bu}.
The expected number of events includes the contribution from signal and
background so that schematically $F_i= \alpha_j F_i^{\rm signal} + \alpha_{j+1}
F_i^{\rm bck}$.  The likelihood (\ref{likelihood}) will be used to calculate, at
a given confidence level, the fraction of values of $\delta_{\rm CP}$ that are
not compatible with the assumed input values.
For T2K, we use 23 energy bins of 50 MeV and for NO$\nu$A 20 bins of 150
MeV. 
In both cases we assume
$\sigma_{13}=0.005$, and all $\sigma_j=0.1$.

\subsection{Relationship between CP exclusion fraction and the uncertainty 
on $\delta_{\rm CP}$}
\label{measures}

An another global measure to display the CP sensitivity used
in the literature is simply to evaluate the uncertainty on the
determination of $\delta_{\rm CP}$, at a certain CL, as a function of
the true value of $\delta_{\rm CP}$. This measure has been used
recently, e.g., in Ref.~\cite{Coloma:2012wq}. We note here that our CP
exclusion fraction has intimate relationship with the uncertainty on 
$\delta_{\rm CP}$ determination.  Since $1 - f_{\rm CPX}$ is equal to the fraction
of the allowed range of $\delta_{\rm CP}$, at least naively, $\left( 1
- f_{\rm CPX} \right) / 2$ would imply the uncertainty associated with
$\delta_{\rm CP} / 2\pi$. Unfortunately this is not quite true, if the
allowed range of $\delta_{\rm CP}$ is disconnected or there are
multiple fake solutions,\footnote{
A related detailed discussion about this point can be found in \cite{Huber:2004gg}.
} 
%
in which case the interpretation can be misleading. In fact, in T2K and
NO$\nu$A with $\sim 10$ years perspective we expect, mainly due to the
lack of statistics, that the determination of $\delta_{\rm CP}$ will
be plagued by large uncertainties and degeneracies, which entails
severe non-Gaussian features of the $\chi^2$ distribution.

On the other hand, in precision measurement era in which $\delta_{\rm
  CP}$ can be measured with high precision, the $\chi^2$ will become
locally Gaussian. Because of the above mentioned properties, we expect
that $\left( 1 - f_{\rm CPX} \right) / 2$ will turn smoothly to be the
uncertainty on $\delta_{\rm CP} / 2\pi$. Therefore, while we prefer to
work with the CP exclusion fraction for the time being because it is
more tolerant to degeneracies, the two measures will become more
closely related to each other in the era of dedicated CP experiments.

\section{Analysis Method} 
\label{method}


In order to simulate T2K $\nu_\mu\to\nu_e$ and
$\bar\nu_\mu\to\bar\nu_e$ events, we used a similar machinery as the
one developed in Ref.~\cite{Machado:2011ar}. We took the fluxes for
the neutrino and antineutrino modes, as well as the backgrounds, from
the Hyper-Kamiokande letter of intent~\cite{Hyper-K}, normalizing the
numbers to the T2K experimental parameters.  We used the cross
sections from Ref.~\cite{Hiraide:2006vh}. The migration of events were
taken into account as below, in a similar way as done
in~\cite{Machado:2011ar}.  We considered four systematic
uncertainties, that is, the signal and background absolute
normalizations for both neutrino and antineutrino modes. We took all
of them to be 10\%, as described in the previous section. In view of
the fact that T2K comes already very close to 10\% level systematic
errors, it is a conservative choice for the neutrino mode, but may be
a reasonable choice for the antineutrino mode.

To mimic the T2K neutrino energy reconstruction, we built migration
matrices for quasi-elastic (QE) and non-quasi-elastic (nQE) events for 
both neutrino and antineutrino modes. For simplicity, we used a 
Gaussian smearing for all cases. For each migration matrix, we set a 
width of the Gaussian and an energy shift at 0.55~GeV and 0.75~GeV,
and we inter/extrapolated for all energies of interest. The precise
values we used are shown in table~\ref{table:t2k}.
\begin{table}
\begin{centering}
\begin{tabular}{|c|c|c|c|c|}
\hline 
 & \multicolumn{2}{c|}{0.55 GeV} & \multicolumn{2}{c|}{0.75 GeV}\tabularnewline
\hline 
 & width (MeV) & shift (MeV) & width (MeV) & shift (MeV)\tabularnewline
\hline
\hline 
$\nu$ QE & 85 & -10 & 98 & -15\tabularnewline
\hline 
$\nu$ nQE & 70 & -325 & 110 & -390\tabularnewline
\hline 
$\overline{\nu}$ QE & 57 & -20 & 60 & -20\tabularnewline
\hline 
$\overline{\nu}$ nQE & 100 & -270 & 120 & -310\tabularnewline
\hline
\end{tabular}
\par\end{centering}
\caption{T2K energy reconstruction parameters used in this paper.}
\label{table:t2k}
\end{table}
The efficiencies were taken to be almost constant for QE events,
around 80\%, and slightly decreasing for nQE, around 25\% and 45\% for
the neutrino and the antineutrino channels, respectively.
We simulate T2K disappearance modes according to
Ref.~\cite{Hiraide:2006vh}, obtaining a sensitivity to $
\sin^22\theta_{23}$ around 0.02 (0.013) at 90\% CL for a 5 (10) years
running only in the neutrino mode.

As a small variation of the value $\vert\Delta m^2_{31}\vert$ within
the current uncertainty (which will be further reduced by
T2K/NO$\nu$A) has little impact on the appearance channel we do not
let it vary in our simulations. We fix it to $2.47\times 10^{-3}$
eV$^2$ ($2.43\times 10^{-3}$ eV$^2$) for the normal (inverted)
hierarchy~\cite{GonzalezGarcia:2012sz}.
For the same reason, we fix the solar neutrino oscillation parameters
to $\sin^2 \theta_{12} = 0.31$ and $\Delta m^2_{21} = 7.54 \times
10^{-5}$ eV$^2$.

Regarding NO$\nu$A simulation, we have based it on the simulation done
in~\cite{Huber:2009xx,Yang_2004}, considering both the appearance and
disappearance channels for the neutrino and antineutrino modes, using
the latest experimental configuration~\cite{Ayres:2007tu,Messier2}.  We used
the fluxes available from~\cite{Messier} and take the cross sections
from Refs.~\cite{Messier:1999kj,Paschos:2001np}. In this case, we do
not put a prior on $\sin^2 2\theta_{23}$, but let it be determined by
NO$\nu$A itself.

Implementing the precisely measured value of $\theta_{13}$ is an
indispensable ingredient in our method of detecting CP violation by
ongoing and near future accelerator experiments
\cite{Minakata:2003wq}.  To incorporate the precision reactor
measurement of $\theta_{13}$, we assume the final sensitivity to match
Daya Bay's current systematic uncertainty of $\sin^2 2\theta_{13}$,
that is, $\delta(\sin^2
2\theta_{13})=0.005$~\cite{An:2012eh,An:2012bu}.

To study the maximum capacity of the experiment to contribute to our
knowledge on the true value of $\delta_{\rm CP}$ we calculate the
fraction of $\delta_{\rm CP}$ values that can be ruled out with a
certain confidence level by the experimental data as a function of the
input parameters $(\sin^2 \theta_{23}^{\rm in}, \delta_{\rm CP}^{\rm
  in})$, either by assuming a known neutrino mass hierarchy or by
marginalizing over it. We have done this for different number of
running years in neutrino and antineutrino modes, as we will describe
in what follows.  We assume that 1 year running of T2K and NO$\nu$A
corresponds, respectively, to delivery of $10^{21}$ and $6 \times
10^{20}$ protons on target (POT). The fiducial mass of
Super-Kamiokande is taken as 22.5 kt and NO$\nu$A detector as 14 kt.

\section{Qualitative discussions based on Bi-probability plots} 
\label{sec:bi-p}

\begin{figure}[htbp]
\begin{center}
\vglue -1.0cm
\hglue -6.5cm 
\includegraphics[width=0.55\textwidth]{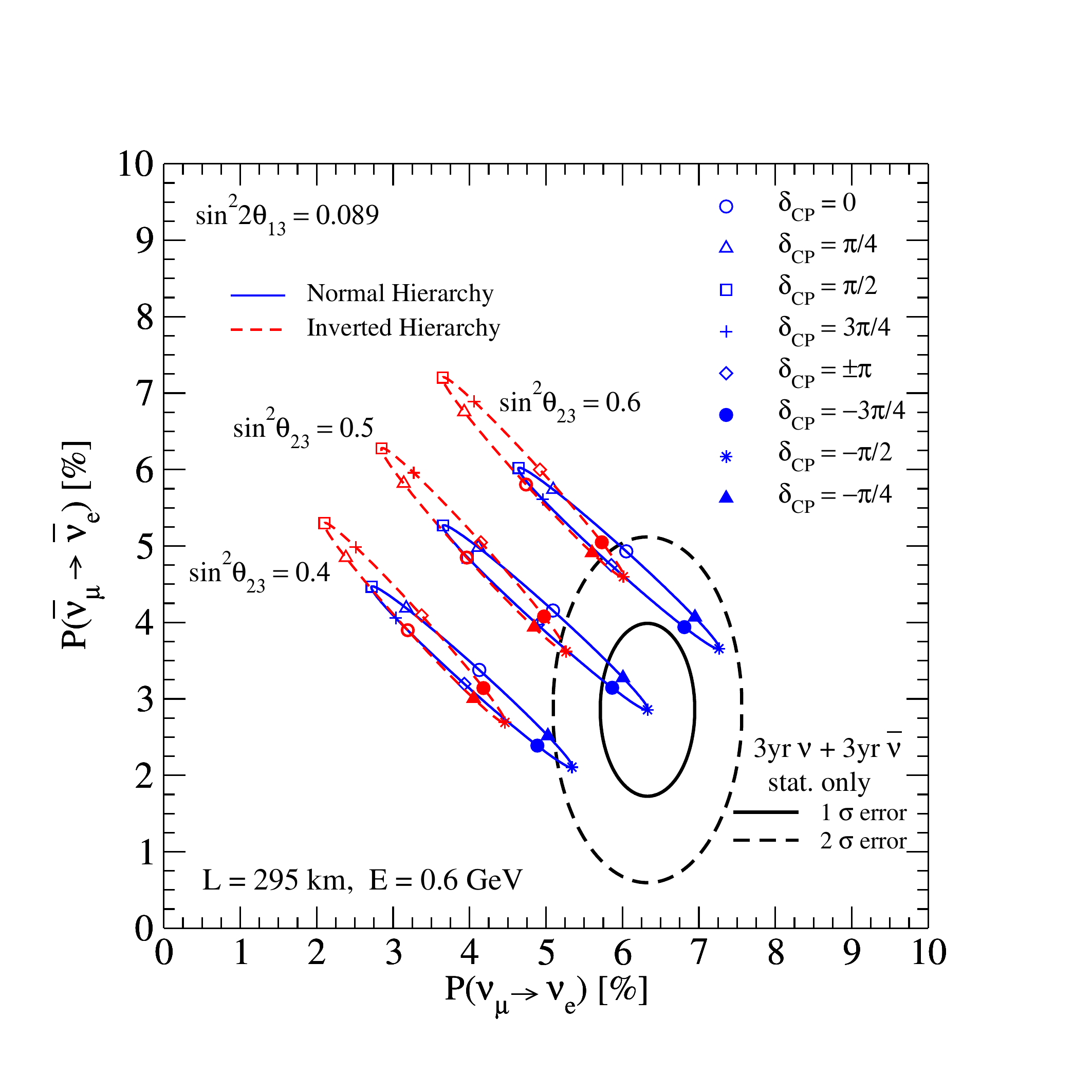} 
\vglue -8.35cm 
\hglue 7.5cm 
\includegraphics[width=0.55\textwidth]{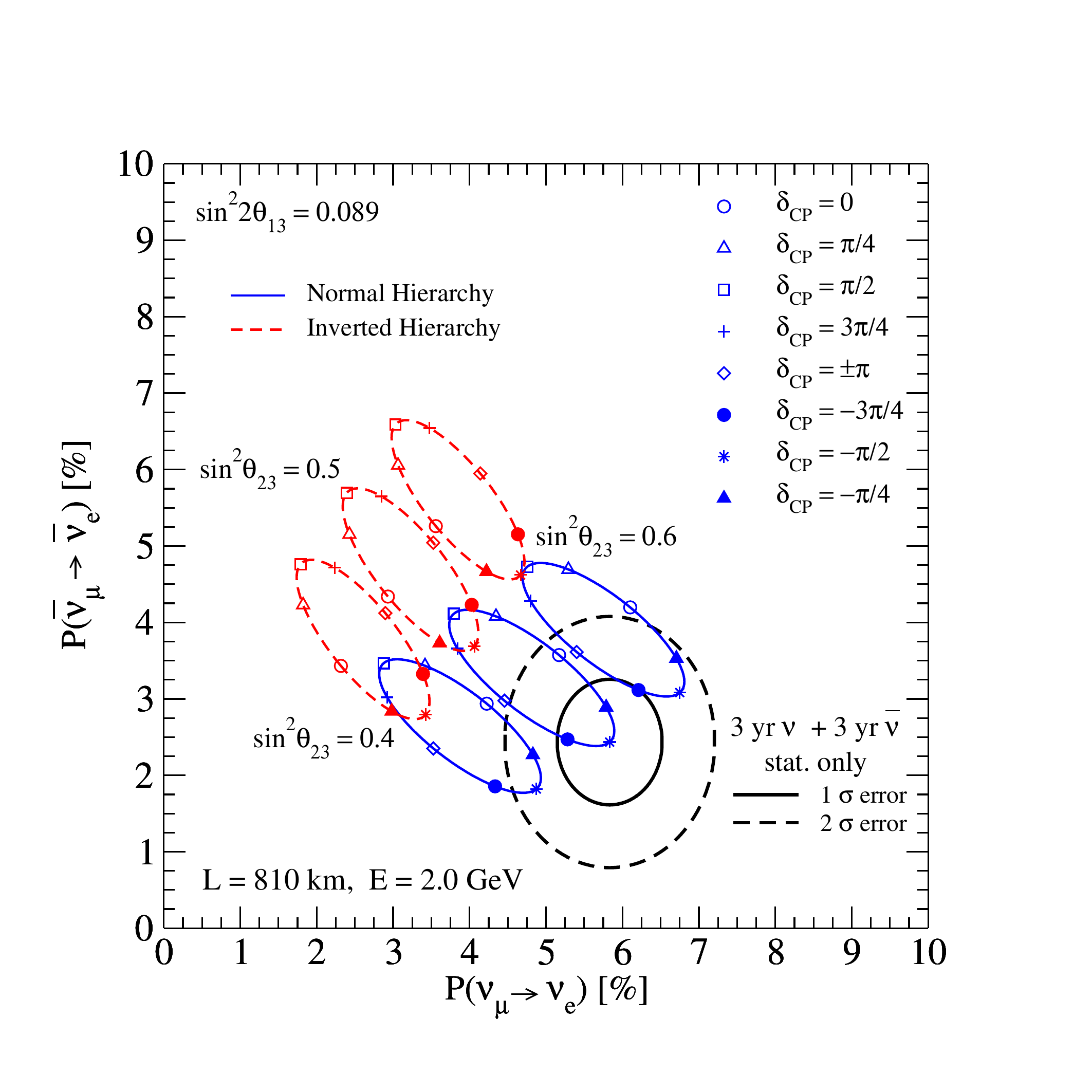}
\end{center}
\vspace{-10mm}
\caption{ Bi-probability plots, or the simultaneous presentation of
  the appearance probabilities, $P(\nu_\mu \to \nu_e)$ and
  $P(\bar{\nu}_\mu \to \bar{\nu}_e)$, by continuously varying
  $\delta_{\text{CP}}$ from $-\pi$ to $\pi$ for the T2K set up of
  $L=295$ km and $E$ = 0.6 GeV (left panel) and the NO$\nu$A one of
  $L=810$ km and $E$ = 2.0 GeV (right panel).  The three ellipses for
  the both mass hierarchies are for $\sin^2 \theta_{23} = 0.4, 0.5,$
  and 0.6 .
In order to have some idea about the precision of the measurements, 
the expected uncertainties for T2K and NO$\nu$A are 
indicated by the solid ($1 \sigma$) and 
dashed ($2 \sigma$) black curves by taking into account
only statistical uncertainties for 3 + 3 years of exposure for
neutrino plus antineutrino modes 
for the case where the mass hierarchy is normal, 
$\delta_{\text{CP}} = -\pi/2$  and $\sin^2 \theta_{23} = 0.5$. }
\label{Bi-P-plots}
\end{figure}

In this section, we present a simple way to understand some of the
notable features in the analysis results presented in
Secs.~\ref{CP-T2K} and \ref{CP-NOVA} by using the bi-probability plots
introduced in Ref.~\cite{Minakata:2001qm}.
As shown in Fig.~\ref{Bi-P-plots} it is a simultaneous presentation of the
appearance probabilities, $P(\nu_\mu \to \nu_e)$ and $P(\bar{\nu}_\mu \to
\bar{\nu}_e)$, calculated by continuously varying $\delta_{\text{CP}}$ from
$-\pi$ to $\pi$ while the other oscillation parameters are fixed. We show in the
left and right panels of Fig.~\ref{Bi-P-plots}, the bi-probability plots which
correspond roughly to the T2K ($L=295$ km and $E$ = 0.6 GeV) and the NO$\nu$A
($L=810$ km and $E$ = 2.0 GeV) setups, respectively, for $\sin^2 \theta_{23} =
0.4, 0.5,$ and 0.6 for both mass hierarchies.

In order to have some idea about the expected precision of the
measurements in terms of the probabilities, we also show in
Fig.~\ref{Bi-P-plots} the expected uncertainty ellipses for the case
where the mass hierarchy is normal, $\delta_{\text{CP}} = -\pi/2$ and
$\sin^2 \theta_{23} = 0.5$, based only on the statistical uncertainty
on the number of signal events assuming the data taking of 3 years
each for neutrinos and antineutrino runs for both NO$\nu$A and
T2K. Note that in reality, the precise evaluation of the uncertainties
is much more complicated as one should take into account several
factors such as energy dependence, backgrounds, systematic
uncertainties and their correlations.

By looking into Fig.~\ref{Bi-P-plots}, one can notice the following
features of the bi-probability plots for T2K and NO$\nu$A setup, which
are very important to understand the results of our analysis shown in
this paper.  \vglue 0.4cm
\noindent
(i) For a given set of oscillation parameters, the CP ellipses for T2K
are thinner and their major axis, which are proportional to the
$\sin\delta_{\text{CP}}$ term in the probability, are longer than that
for NO$\nu$A. These properties follow because the neutrino energy
taken for T2K is closer to the first oscillation maximum, $|\Delta
m^2_{32}|L/(4E) = \pi/2$.

\vglue 0.2cm
\noindent
(ii) For a given set of oscillation parameters, the two CP ellipses
for the normal and the inverted mass hierarchies are more separated
for NO$\nu$A than for T2K due to a stronger matter effect in the
former setup.  \vglue 0.4cm
\noindent
From these observations one can naively expect that the feature described in (i)
would make T2K more sensitive than NO$\nu$A to $\delta_{\text{CP}}$
determination, assuming that the numbers of events for these two experiments are
similar. While the one in (ii) is the feature familiar to us, it would make
NO$\nu$A more sensitive than T2K to the mass hierarchy, which potentially could
help also in increasing the sensitivity to $\delta_{\text{CP}}$ by reducing the
degeneracy related to the unknown mass hierarchy.

Let us discuss the importance of exploiting the observation of both
the neutrino and the antineutrino modes.
For the purpose of illustration, let us look at the bi-probability
plot for the T2K experiment. 
Let us consider the case of the normal mass hierarchy.
Suppose that $\sin^2 2\theta_{23} = 0.96$
(corresponds to $\sin^2\theta_{23} = 0.4$ or 0.6), and that only the
appearance probability in the neutrino mode is observed with the
result $P(\nu_\mu \to \nu_e) = 5\%$.
Then, we can not distinguish the cases between $\sin^2 \theta_{23} =
0.4$ with $ -3\pi/4 \lsim \delta_{\text{CP}} \lsim -\pi/4$, and
$\sin^2 \theta_{23} = 0.6$ with $\pi/4 \lsim \delta_{\text{CP}} \lsim
3\pi/4$ (see the left panel of Fig.~\ref{Bi-P-plots}).
However, if the antineutrino appearance probability is also measured, these two
cases can be distinguished since their probabilities are rather different,
$P(\bar{\nu}_\mu \to \bar{\nu}_e) \sim 2\%$ for $\sin^2 \theta_{23} = 0.4$
against $P(\bar{\nu}_\mu \to \bar{\nu}_e) \sim 6\%$ for $\sin^2 \theta_{23} =
0.6$, and consequently, one can constrain better also the allowed range of
$\delta_{\text{CP}}$. Therefore, the additional running of the $\bar{\nu}_\mu
\to \bar{\nu}_e$ mode clearly helps. This is true also for the NO$\nu$A
experiment, as the qualitative behavior of the CP ellipses for NO$\nu$A is
similar to that for T2K, see the right panel of Fig.~\ref{Bi-P-plots}.

Let us also comment about some expectations on CP sensitivity to
  establish CP violation. In Fig.~\ref{Bi-P-plots}, we assume that we
  know the true mass hierarchy and consider only the statistical
  uncertainties. Despite the optimistic assumption it is clear that we
  can not establish CP violation at 3$\sigma$ CL by either one of these 
  experiments (T2K or NO$\nu$A), or by both, after 3 + 3 years of $\nu$
  and $\bar{\nu}$ running. Therefore, the presence or absence of CP
  violation may not be a useful measure for these
  experiments. However, even in the case without capability of
  establishing CP violation, depending on the true value of
  $\delta_{\text{CP}}$, it may be possible to exclude a certain range
  of $\delta_{\text{CP}}$ values at some CL.  
Therefore, it can be useful to quantify the sensitivity of experiments
to measure $\delta_{\rm CP}$, by using the CP exclusion fraction, as
discussed in this paper.  From Fig.~\ref{Bi-P-plots}, one expects that
T2K can exclude larger ranges of $\delta_{\text{CP}}$ than NO$\nu$A,
for the same exposure.  This expectation was confirmed by the actual
calculation in in Sec. \ref{CP-NOVA}.
\begin{figure}[htbp]
\vspace{-0.5cm}
\begin{center}
\vspace{-1mm}
\includegraphics[width=0.6\textwidth]{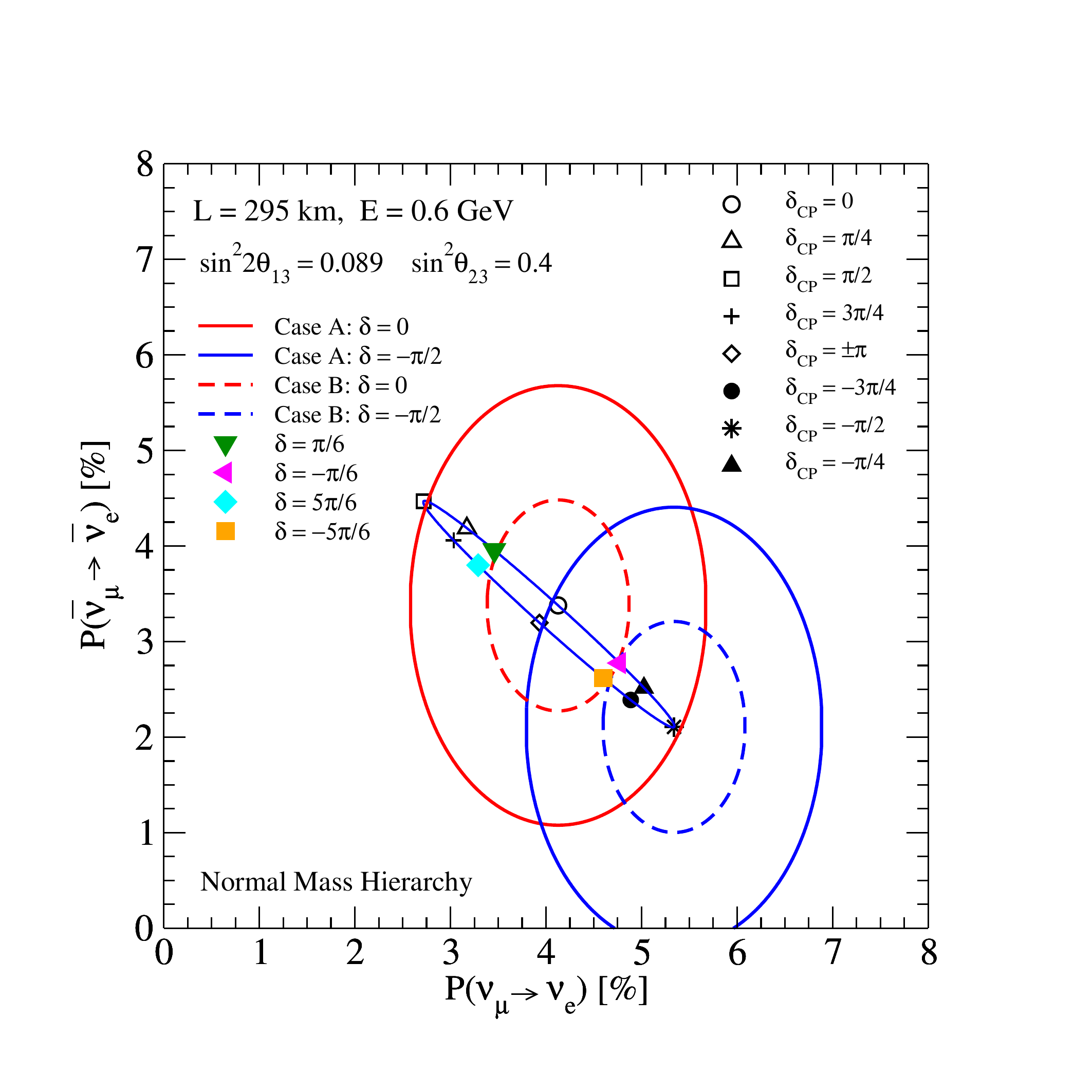}
\end{center}
\vspace{-10mm}
\caption{ Bi-probability plot for T2K for $\sin^2 \theta_{23} = 0.4$
  and the normal mass hierarchy with the error ellipse for
  $\delta_{\text{CP}} = 0$ and $-\pi/2$, which explains the impact of
  the increase of statistics on the CP exclusion fraction.  }
\label{Bi-P-plots-2}
\end{figure}

What would be the values of $\delta_{\text{CP}}$ which give larger or 
smaller CP exclusion fractions? 
The answer depends on the statistics 
as well as our knowledge on the mass hierarchy.
Let us look at
Fig.~\ref{Bi-P-plots-2} in which the bi-probability plot with the normal mass
hierarchy is shown for $\sin^2 \theta_{23} = 0.4$.
Suppose that the true value of $\delta_{\text{CP}}$ is zero and the
statistics is so low that the error ellipse cover the whole range of
$\delta_{\text{CP}}$ as the red solid curve in Fig.~\ref{Bi-P-plots-2}
does. In this case, no region of $\delta_{\text{CP}}$ space would be
excluded. Even in this case, however, it is possible to exclude
roughly half of the range of $\delta_{\text{CP}}$ if the true value of
$\delta_{\text{CP}}$ is equal to $\pm \pi/2$, as we can see from the
solid blue curve ($\delta_{\text{CP}} = - \pi/2$) in
Fig.~\ref{Bi-P-plots-2}.  Therefore, $\delta_{\text{CP}} = \pm \pi/2$
is the most favorable value for highest CP exclusion fraction whereas
$\delta_{\text{CP}} = 0$ is the least.  

The situation would change significantly as the statistics increases.
If we compare the cases of $\delta_{\text{CP}} = 0$, the red dashed
curve, and $\delta_{\text{CP}} = -\pi/2$, the blue dashed curve in
Fig.~\ref{Bi-P-plots-2} we observe that the exclusion fraction of
$\delta_{\text{CP}}$ region of these two cases become comparable due
to a Jacobian effect, and is approximately equal to $2/3$. If the
statistics increases further, the exclusion fraction for
$\delta_{\text{CP}} = 0$ is expected to be larger than that for
$\delta_{\text{CP}} = -\pi/2$. Thus, the favorable and unfavorable
values of $\delta_{\text{CP}}$ for the CP exclusion fraction will be
interchanged as the statistics increases.
This can be confirmed by our results shown in 
Secs.~\ref{CP-T2K} and \ref{CP-NOVA}, 
by comparing, for e.g., the right top panel of Fig.~\ref{T2K-5}
for T2K 2+3 running and the middle right panel 
of Fig.~\ref{T2K-NOVA} for T2K 10 + 10 running where the mass 
hierarchy was assumed to be known. 
Even if the hierarchy is unknown, the same feature can be seen 
when T2K and NO$\nu$A are combined as the hierarchy 
information comes from the result of the fit in this case. 

\begin{acknowledgments}
We are indebted to Masashi Yokoyama for his numerous suggestions and
help kindly offered to us to improve our analysis of the T2K
experiment.
H.M. thanks CNPq for support for Professor Visitante to the
Departamento de F\'{\i}sica, Pontif{\'\i}cia Universidade Cat{\'o}lica
do Rio de Janeiro. He is also partially supported by KAKENHI received
through Tokyo Metropolitan University, Grant-in-Aid for Scientific
Research No. 23540315, Japan Society for the Promotion of Science.
This work was supported by Funda\c{c}\~ao de Amparo \`a Pesquisa do
Estado de S\~ao Paulo (FAPESP), Funda\c{c}\~ao de Amparo \`a Pesquisa
do Estado do Rio de Janeiro (FAPERJ), Conselho Nacional de Ci\^encia e
Tecnologia (CNPq).  R.Z.F. and P.A.N.M. also thank Institut de
Physique Th\'eorique of CEA-Saclay for the hospitality during the time
this work was developed and acknowledge partial support from the
European Union FP7 ITN INVISIBLES (PITN-GA-2011-289442).
\end{acknowledgments}

\end{document}